\documentclass[submission, Phys]{SciPost}

\usepackage[utf8]{inputenc}
\usepackage{amsmath,amssymb,amsfonts,graphicx,cite,multirow,tikz,enumitem}
\usepackage[style=base]{caption}
\usepackage{pstricks,pdflscape,morefloats,algorithmicx,algpseudocode,slashed}
\usepackage[section]{algorithm}
\usepackage{booktabs}

\usepackage{xspace}
\usepackage{comment}

\makeatother

\usepackage{color}
\usepackage{hyperref}
\usepackage{subcaption}

\def\qb{\bar q}

\newcommand{\pT}{\ensuremath{p_{\perp}}\xspace}

\usepackage{mathtools}

\author{Cody~B~Duncan and Peter~Skands}

\begin{document}

\begin{center}
    {\Large \textbf{Fragmentation of Two Repelling Lund Strings}}
\end{center}

\begin{center}
    Cody~B~Duncan\textsuperscript{1*} and Peter Skands\textsuperscript{1}
\end{center}

\begin{center}
    \textbf{1} School of Physics \& Astronomy, Monash University, Clayton, VIC 3800, Australia\\[1ex]
    * cody.duncan@monash.edu
\end{center}

\renewcommand{\abstract}[1]{\section*{}\vspace{-45pt}\textbf{#1}}
\abstract{Motivated by recent discoveries of flow-like effects in pp collisions, and noting that multiple string systems can form and hadronize simultaneously in such collisions, we develop a simple model for the repulsive interaction between two Lund strings with a positive (colour-oriented) overlap in rapidity. The model is formulated in momentum space and is based on a postulate of a constant net transverse momentum being acquired per unit of overlap along a common rapidity direction. To conserve energy, the strings shrink in the longitudinal direction, essentially converting a portion of the string invariant mass $m^2$ into $p_\perp^2$ for constant $m_\perp^2 = m^2 + p_\perp^2$ for each string. 
The reduction in string invariant mass implies a reduced overall multiplicity of produced hadrons; the increase in $p_\perp^2$ is local and only affects  hadrons in the overlapping region.
Starting from the simplest case of two symmetric and parallel strings with massless endpoints, we generalize to progressively more complicated configurations. We present an implementation of this model in the Pythia event generator and use it to illustrate the effects 
on hadron $p_\perp$ distributions and dihadron azimuthal correlations, contrasting it with the current version of the ``shoving'' model implemented in the same generator.}

\vspace{10pt}
\noindent\rule{\textwidth}{1pt}
\tableofcontents\thispagestyle{fancy}

\section{Introduction}
\label{sec:introduction}
Hadronization models play an essential role in the description of hadronic events in high-energy collisions, connecting the short-distance physics of quarks and gluons with the observable world of colourless 
(long-lived) hadrons via a dynamical process that enforces confinement. 
The two major models of hadronization
used in proton-proton event generation are the Lund string model~\cite{Andersson:1983ia,Andersson:1983jt,Andersson:1998tv,SJOSTRAND1984469} and the
cluster model~\cite{FIELD198365,GOTTSCHALK1984349,WEBBER1984492}, with the former implemented
in Pythia~\cite{Sjostrand:2006za,Sjostrand:2007gs,Sjostrand:2014zea} and Epos~\cite{Werner:2005jf,Pierog:2013ria}, and the latter in Herwig
\cite{Bahr:2008pv,Bellm:2015jjp,Bellm:2017bvx} and Sherpa \cite{Gleisberg:2008ta,Bothmann:2019yzt}. 

While the Lund string model has been able to qualitatively describe
a large number of hadron-level observables from $e^+e^-$ to proton-proton collisions across a wide range of CM energies (see e.g.~\textsc{mcplots}~\cite{Karneyeu:2013aha}), 
recent data in particular from the LHC experiments have highlighted some  shortcomings. ALICE has shown unequivocally that strangeness production increases as a function of event multiplicity in minimum-bias event samples~\cite{Adam:2015qaa,Acharya:2018orn}, 
while CMS discovered the near-side ridge in high-multiplicity events~\cite{Khachatryan:2010gv,Velicanu:2011zz}. The latter has  been elaborated upon in a number of studies by both ATLAS and CMS~\cite{Aad:2015gqa,Khachatryan:2015lva,Khachatryan:2016txc,Aaboud:2016yar,Aaboud:2017acw}, and there are also several additional  indications of strangeness enhancement e.g.\ in the underlying event~\cite{Chatrchyan:2013qsa,Aad:2019xek,Cui:2019jbt}. 
Both of these phenomena are widely believed to have their roots in collective effects, but
in the baseline Lund string model, 
each string hadronizes independently of the others 
(modulo effects of colour reconnections, see, e.g.~\cite{Sjostrand:2013cya,Sjostrand:2017ele}).

Several proposals have been made that can potentially explain these phenomena.
Rope hadronization~\cite{Bierlich:2014xba,Bierlich:2015rha} takes aligned strings in rapidity and enhances their string tensions based on a Casimir scaling argument~\cite{AMBJORN1984533,Bali:2000un}, leading to increased strangeness production and higher average $p_\perp$ values in string breaks. Shoving, a mechanism for microscopic string-string interactions which generates transverse momentum pressure between overlapping strings, was
proposed in \cite{Bierlich:2016vgw, Bierlich:2017vhg} and showed long-range azimuthal correlations. 
Both the Rope and shoving model have been implemented
in Pythia, Dipsy~\cite{Flensburg:2011kk}, and Angantyr \cite{Bierlich:2018xfw}. Alternatively, the approach taken by 
Epos~\cite{Pierog:2013ria} invokes the notion of a critical string density beyond which a heavy-ion inspired hydrodynamic modelling takes over, which includes collective flow and thermally enhanced strangeness production. Yet a third line of argument is that colour reconnections (CR) can produce flow-like effects~\cite{Ortiz:2013yxa}, essentially by creating net boosted hadronizing systems. Baryon-to-meson ratios may also be altered by CR effects~\cite{Christiansen:2015yqa} but would have to be supplemented by something like rope hadronization to significantly alter net strangeness fractions. 

In~\cite{Fischer:2016zzs}, the authors studied the effects of a thermodynamic string fragmentation model, which used an exponential transverse mass spectrum instead of the usual Gaussian form for the Lund string model.
Recent work on the cluster model has also tried to capture some
of the collective-like effects seen by introducing 
baryonic clusters and strangeness enhancement in Herwig
\cite{Gieseke:2018gff,Duncan:2018gfk}. Other approaches to studying and modeling azimuthal correlations in proton-proton collision environments include the String Percolation model~\cite{Celik:1980td,Ferreiro:2003dw,Braun:2015eoa,Bautista:2015kwa,Ramirez:2017oef}, or interference effects from multiple parton interactions~\cite{Blok:2018xes} or from the BFKL parton shower evolution process~\cite{Levin:2011fb}. Recent work has investigated the initial state geometry of the collision and the resultant effect on azimuthal anisotropy~\cite{Bzdak:2013zma,Yan:2014nsa,Albacete:2016gxu}. The Colour Glass Condesate (CGC), a successful framework for describing the collision environment in heavy-ion collisions, has also been applied to proton-proton collisions~\cite{Dumitru:2010iy,Dusling:2012iga}. Kinetic transport theory has also been used to study the potential source of angular correlations~\cite{Kurkela:2018qeb}. A review of collectivity in small systems can be found in e.g.~\cite{Schlichting:2016sqo,Bierlich:2020kzy}.

We here take the same basic starting point as the shoving model~\cite{Bierlich:2016vgw, Bierlich:2017vhg}, namely that nearby Lund strings should exert a force upon one another. We focus  on repulsive forces since we assume that colour-reconnection models such as~\cite{Christiansen:2015yqa} (based on colour algebra and string-length minimisation) provide a first approximation to any attractive effects.  
We further assume that all of the hadronizing colour charges emanate from a region that is small compared with the typical width of a string. This restricts the applicability of our model to small systems but allows us the simplification of working entirely in momentum space. By contrast, the shoving model adopts an explicit picture of the spatial distribution and time evolution of the strings. (The space-time structure of hadronization in the Lund model was also recently further explored in~\cite{Ferreres-Sole:2018vgo}.)
Furthermore, the effect of the interaction is in our model represented via a global rescaling of the 4-momenta of the string endpoints combined with a local addition of $p_\perp$ to hadrons formed in regions of string overlap, while the shoving model imparts transverse momentum by adding a number of low-energy slightly massive gluons to each string. Despite similar physical starting points, we therefore do expect some qualitative differences to arise between the shoving model~\cite{Bierlich:2016vgw, Bierlich:2017vhg} and our momentum-space realization of repelling strings.

The article is organized as follows. Sec.~\ref{sec:string} presents a short review of the Lund string model with emphasis on those 
features that are most relevant to our toy extension model. 
Sec.~\ref{sec:simplestStrings} introduces our
string-string interaction model in the context of the simplest 
two-string configuration, and presents how the repulsion is implemented
during string fragmentation, and the effects on primary hadron transverse
momentum.
We then extend this formalism to a more general parallel
two-string configuration in Sec.~\ref{sec:generalStrings} and then to strings
with endpoints with both longitudinal and transverse momentum in Sec.~\ref{sec:boosted}. To make a connection with the phenomenological characterisations of collective flow used in heavy-ion inspired studies, we illustrate the effects on two-particle cumulants, $c_2\{2\}$, for selected two-string configurations in Sec.~\ref{sec:cumulant}.
In Sec.~\ref{sec:finalState}, we discuss the effects of decays of short-lived primary hadrons. 
Modifications for strings with massive endpoints are briefly discussed in Sec.~\ref{sec:massive} before we conclude and give an outlook for future work in Sec.~\ref{sec:conclusion}.

\section{Lund String Model}
\label{sec:string}
The Lund string model
\cite{Andersson:1983ia,Andersson:1983jt,SJOSTRAND1984469,Andersson:1998tv} is based on the linear nature of the confinement potential $V(r) = \kappa r$ between static quark-antiquark pairs separated by distances greater than about a femtometre (see e.g.~\cite{Bali:2000un}). 
Strings are implemented in Pythia at the end of the perturbative shower, where long colour-chains produced by the shower are collected
into colour singlets, the so-called Lund strings. 

A Lund string represents a confined gluonic flux tube or vortex line. In the simplest case it runs between a quark endpoint via any number of intermediate gluons (which generate transverse kinks in the structure) to an antiquark endpoint. Other colour topologies are possible as well, such
as junctions and gluon loops. In this work, 
we restrict our attention to simple  
$q\qb$ strings 
without any transverse gluon excitations.

As the endpoints propagate outward in opposite directions from the production point, their energy and momentum gets transferred to
the Lund string that stretches between them. 
When sufficient energy is available, new $q'\bar{q}'$ pairs can be produced in the string field (typically by invoking a Schwinger-type tunneling mechanism~\cite{Schwinger:1951nm}); the string thereby breaks into successively shorter pieces each of which ultimately becomes an on-shell hadron, in a process called fragmentation. In the 
ordinary Lund string model, each string fragments independently, and
each string break is independent of any others. 

Fragmentation proceeds by successively splitting off one hadron from either endpoint (chosen at random), with the created hadron at each step taking a fraction $z$ of the string's available lightcone momentum distributed according to the Lund symmetric fragmentation function:
\begin{equation}
    f(z) = N \frac{(1-z)^a}{z}\exp\left(\frac{-bm_{\perp}^2}{z}\right)
    ~,
    \label{eq:fragmentation}
\end{equation}
with the leftover string retaining the remainder $1-z$. $N$ is a normalisation constant and $a$ and $b$ are phenomenological parameters to be determined from fits to data, see e.g.~\cite{Skands:2014pea,Amoroso:2018qga}.
$m_{\perp}^2 = m^2 + p_\perp^2 $ is the transverse mass of the produced hadron; its $p_\perp$ is obtained as the (vector) sum of the $p_\perp$ values of each of its constituent quarks. In the absence of collective effects each string break is assumed to impart an equal and oppositely oriented $p_\perp$ to the produced quark and antiquark, which by default is given a Gaussian distribution, by analogy with the Schwinger mechanism in QED~\cite{Schwinger:1951nm}. 
In the Rope model~\cite{Bierlich:2014xba}, the coherent fragmentation of multiple nearby colour charges can cause the width of this $p_\perp$ distribution (as well as strangeness and baryon production probabilities) to increase. While we believe those arguments to be fundamentally correct, for simplicity we focus in this work solely on the collective repulsion aspect, 
keeping other string-breaking aspects unmodified. 

\subsection{Fragmentation and rapidity}
\label{sec:rapiditySpan}
In the context of interacting strings, we will be interested in the effective overlap in rapidity between a produced hadron and a nearby string piece. To start with, we need an expression for the rapidity span taken by each hadron along an axis defined by its own string system.

Letting $m_0$ denote a generic hadron mass, the rapidity span of a simple $q\qb$ string with massless endpoints traveling in opposite directions along the $z$-axis is:
\begin{equation}
    \begin{split}
        \Delta y_0 &= \ln\left(\frac{W_{+q}}{m_0}\right) -\left[- \ln\left(\frac{W_{-\bar{q}}}{m_0}\right)\right] ,\\
        &= \ln\left( \frac{W^2}{m^2_0}\right) ,
    \end{split}
    \label{eq:simpleRapSpan}
\end{equation}
where $W_\pm = E\pm p_z$ are their lightcone momenta, and $W^2 = W_+ W_-$ is the squared invariant mass of the string. Throughout this work, we will use the $z$ axis as the (common) rapidity axis, and our example configurations will be defined so that this is reasonable, but there is obviously nothing special about this choice; the formalism we develop can be applied for any choice of axis.  

After a hadron, $h$, is split off from one of the endpoints, let the invariant mass of the leftover string be $W'^2$. The size of the rapidity interval associated with the produced hadron can then be identified with the difference:
\begin{equation}
    \Delta y_h = \ln\left( \frac{W^2}{m_0^2} \right) - \ln\left( \frac{W'^2}{m_0^2} \right) =\ln\left( \frac{W^2}{W'^2} \right)~,
    \label{eq:simpleRap}
\end{equation}
which is independent of $m_0$. 
App.~\ref{app:fragmentationDetails} elaborates on how Eq.~(\ref{eq:simpleRap}) relates to the sequence of $z$ fractions and hadron mass values for arbitrary (sequences of) string breaks, using the notation from \cite{SJOSTRAND1984469,Ferreres-Sole:2018vgo} which also matches the
code implementation. Below, we shall use these expressions to quantify the total rapidity overlap that a given hadron has with a nearby string piece.  

\section{Repulsion Between Two Parallel Identical Strings}
\label{sec:simplestStrings}
We  start by considering the simplest possible configuration: two straight and parallel strings of the same squared invariant mass, $W^2$. 

Viewed in space-time, the repulsion between two such strings should depend on their (time-dependent) transverse separation distance~\cite{PhysRevB.3.3821,PhysRevB.83.054516}. However, in the context of hadronization in high-energy particle collisions, the preceding perturbative stages of event generation are normally treated in momentum space, i.e.\ in terms of plane-wave approximations that are not well localized in space-time. Thus, one faces a problem of mapping partons represented in momentum space onto string systems represented in space-time. In the framework of classical string theory, on which the Lund model is based, one may simply use the string tension $\kappa$ to convert between the two pictures. 
But when multiple string systems are involved, any interactions between them will depend on the space-time separation between the production points of each system, which the momentum-space perturbative boundary conditions only serve to fix up to an ambiguity $\propto 1/\Lambda_\mathrm{QCD}$. 
Moreover, while a strict classical interpretation would in principle allow for arbitrarily small separations, string descriptions are only appropriate for long-distance QCD. 
Interesting work has been done recently  to bridge the two pictures~\cite{Ferreres-Sole:2018vgo,Bellm:2019wrh}, but for the purpose of this study we would like to explore how far we can get if we stay in momentum space. 

Our underlying assumption will be that our colliding systems are of order a hadronic size (hence we do not address heavy ions) and that, by the time strings are formed, they are already at least some ``typical'' transverse distance apart, again of order hadronic sizes even if the directions of motion of the endpoints were originally completely parallel. We make the boost-invariant ansatz that parallel strings impart a constant amount of net transverse momentum to each other per unit of overlap in rapidity,
\begin{equation}
    \frac{\mathrm{d}p_{\perp R}}{dy} = c_R~,\label{eq:dptdy}
\end{equation}
where the constant $c_R$, which has dimensions of GeV per unit rapidity, represents the main tuneable parameter in our model. It controls the strength of the repulsion, or alternatively, the conversion strength of longitudinal momentum into transverse momentum.\footnote{In a future extension we shall relate this to an increase in the tension of the individual strings as well, in a manner similar to what is done in Rope hadronization, but this is outside the scope of this
work.}
Non-parallel configurations will be discussed below. 
We further make the ansatz that each hadron produced in the overlap region receives a fraction of the total repulsion $p_\perp$ in proportion to (the overlapping portion of) its rapidity span according to 
Eq.~\ref{eq:simpleRap}. 

\begin{figure*}[tp]
\centering
\includegraphics[width=0.95\textwidth]{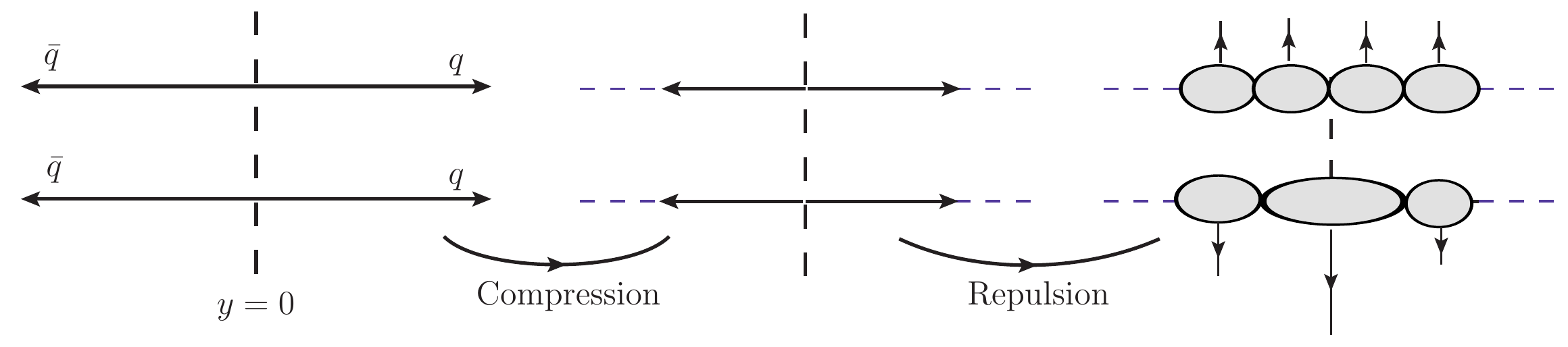}

\caption{ Schematic diagram of the simplest two-string configuration
and the two steps in our model: compressing the strings (black solid lines) and repulsion during
the string's fragmentation. The hadrons (grey ovals) receive \pT proportional to the
string length they take. We have ignored the Gaussian transverse momentum
generation in the Lund string model for the purposes of this figure.
The transverse separation between the strings in the diagram is for clarity.}
\label{fig:simpleStrings}
\end{figure*}

A schematic diagram of how our model works is shown in Fig.~
\ref{fig:simpleStrings}. In a first step, we remove an amount of longitudinal momentum from the original endpoints (``compression''), in proportion to the size of the total rapidity overlap between the two strings. 
In the second step, the energy that was removed in the compression step is imparted back to the hadrons formed in the region(s) of overlap, as transverse momentum (``repulsion'').

\subsection{String Compression}
\label{sec:simpleCompression}
Each string is defined by its two endpoints, which for simplicity we take to be massless for now and travelling in opposite directions along the $z$ axis. Right-moving endpoints thus start out with  lightcone momenta $W_{+} = 2E$  and $W_- = 0$ and vice versa for the left-moving ones. 
In the fully symmetric setup we consider 
here, both strings will undergo the same transformations described below. 
We focus on just one of them. 

Since the strings have equal invariant masses, the overlap is simply
the full rapidity span of each string, i.e. 
$\Delta y_{\mathrm{ov}} = \Delta y_{\mathrm{string}}$, which is given by
Eq.~(\ref{eq:simpleRapSpan}). For this work, we found that using
too small an $m_0$ can lead to pathological results since this presumes
that every hadron you can create has an invariant mass of that order.
Instead, we will choose to work with $m_0 = m_{\rho} = 0.77$~GeV.
Thus, by integration of Eq.~(\ref{eq:dptdy}) the \pT gained by each string will be:
\begin{equation}
    p_{\perp R} = \pm c_{R} \cdot \Delta y_{\mathrm{ov}} ,
    \label{eq:pTrepel}
\end{equation}
where the $\pm$ sign symbolically represents that the kicks will act in opposite directions, so 
that no net $p_\perp$ is gained by the string-string system as a whole. 

To conserve energy, this $p_\perp$ must be acquired at the expense of some amount of 
longitudinal momentum. 
We start by defining a set of intermediate rescaled lightcone momenta $W_{\pm}' = f_{\pm} W_{\pm}$ with 
\begin{equation}
    f_+ f_- = 1 - \frac{p^2_{\perp,R}}{W^2} \leq 1,
    \label{eq:shorteningFactor}
\end{equation}
which corresponds to a $W'$ string system with a lower invariant mass, 
\begin{equation}
W'_- W'_+ = W'^2 = W^2 - p_{\perp R}^2~. 
\end{equation}
This first step of the model is illustrated by the left-hand part of Fig.~\ref{fig:simpleStrings}, labelled ``Compression''.
In the simple case studied in this section the compression factors $f_+$ and $f_-$ must be equal for symmetry reasons. (More general cases, with $f_+ \neq f_-$, will be considered in the next section.)

A particularly simple way of representing the repulsion effect would be to boost 
the $W'$ system transversely by a factor $\vec{\beta}_\perp = \vec{p}_{\perp R}/W'$. However, as G.~Gustafson demonstrated
during enjoyable discussions in Lund, such a boost would assign relatively more of the repulsion $p_{\perp}$ to high-rapidity hadrons than to central ones, in contrast with the manifestly longitudinally invariant form of Eq.~(\ref{eq:dptdy}). 
Instead, we therefore modify the fragmentation of the $W'$ system in a more local way, by allowing each produced hadron to receive an additional amount of $p_{\perp}$ in a manner designed to reproduce Eq.~(\ref{eq:dptdy}). 

Writing the 4-vectors as $\left(p_+,p_-, \vec{p}_{\perp}\right)$, the $W'$ system is defined by:
\begin{equation}
\begin{split}
    p'_{q} & = fW_+\left( 1, 0, \vec{0}_{\perp}\right) , \\
    p'_{\qb} & = fW_-\left( 0, 1, \vec{0}_{\perp}\right)~.
\end{split}
\end{equation}
As remarked above, this has a lower total energy, $W'$, than that of the original system. The ``missing energy'' will gradually be added back during the fragmentation process, in the form of additional $p_\perp$ given to the hadrons that are formed in the region(s) of overlap. Unlike the standard fragmentation $p_\perp$ in string breaks, which is randomly and independently distributed in azimuth for each breakup, 
a single global $\phi$ choice characterises the $p_\perp$ component from repulsion (with $\pi+\phi$ used for the hadrons in the recoiling string system).
We will now discuss the details of this second step, illustrated by the right-hand part of Fig.~\ref{fig:simpleStrings}, labelled ``Repulsion''.

\subsection{Repulsion}
\label{sec:fragmentationRepulsion}
As mentioned in Sec.~\ref{sec:rapiditySpan}, 
we can assign a rapidity
span to each hadron as it gets produced by the rapidity span lost by the string when producing the hadron.
Using Eq.~(\ref{eq:hadronspan}), a hadron receives a corresponding
fraction of $p_{\perp R}$, calculated in the same manner as Eq.~(\ref{eq:pTrepel}):
\begin{equation}
    p_{\perp h} ~=~ c_{R} \Delta y_h ~=~ p_{\perp R} \frac{\Delta y_h}{\Delta y_{\rm string}},
    \label{eq:pThadron}
\end{equation}
where $\Delta y_h$ is the rapidity span of the string taken by the hadron, such
that $\sum \Delta y_h = \Delta y_{\rm string}$, and consequently
the summed repulsion momentum given to hadrons is equal to the total
repulsion momentum. 
Generalising to cases in which the two strings do not fully overlap, the numerator and denominator of the rapidity-span ratio in the last expression can simply be changed to refer to the overlapping portions of the hadron and total rapidity spans, respectively.
After the hadron receives the repulsion $p_\perp$, its energy is then adjusted by the amount required to put it back on shell. In this way, the ``missing energy'' discussed above is gradually added back to the system. 

Note that, if there were no other sources of transverse momentum, putting a hadron on-shell after the repulsion would always increase its energy. However, since each string break is associated with a randomly distributed fragmentation $p_\perp$ (with each hadron in general receiving contributions from two such breaks), which must be added vectorially to the repulsion \pT, some hadrons may have lower total \pT\ after adding the repulsion effect.  In our modeling setup, such hadrons are regarded as donating some energy back to the string system's reservoir of ``missing energy'', with the sum over all hadrons still respecting eq.~(\ref{eq:pTrepel}). 

With this modification, we follow the same iterative fragmentation procedure as in ordinary Pythia, splitting off hadrons from either end,  allowing them to receive additional repulsion $p_\perp$ and putting them back on shell, until the invariant mass of the remaining string system drops below a cutoff value: 
\begin{equation}
    W_{\mathrm{rem}}^2 < W^2_{\mathrm{stop}} .
\end{equation}
At this point, we add any remaining repulsion \pT to the remnant object,  as well as any energy that is still missing from the compression process. This makes total energy and momentum conservation explicit. 
Pythia then produces two final hadrons from this modified remnant string.

\subsection{Results}
\label{sec:simpleResults}

\begin{figure*}[tp]
\centering
\includegraphics[width=0.49\textwidth]{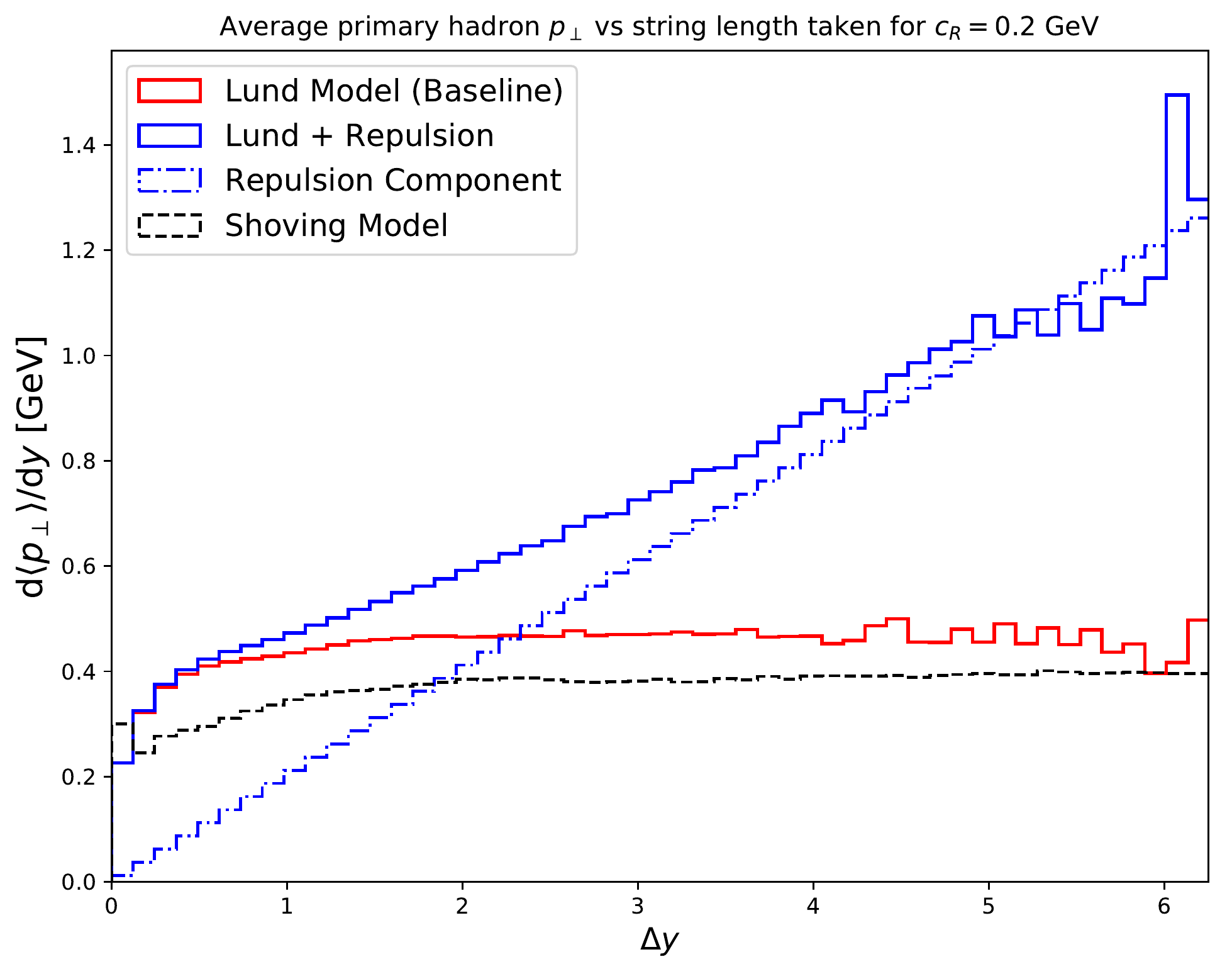}
\includegraphics[width=0.49\textwidth]{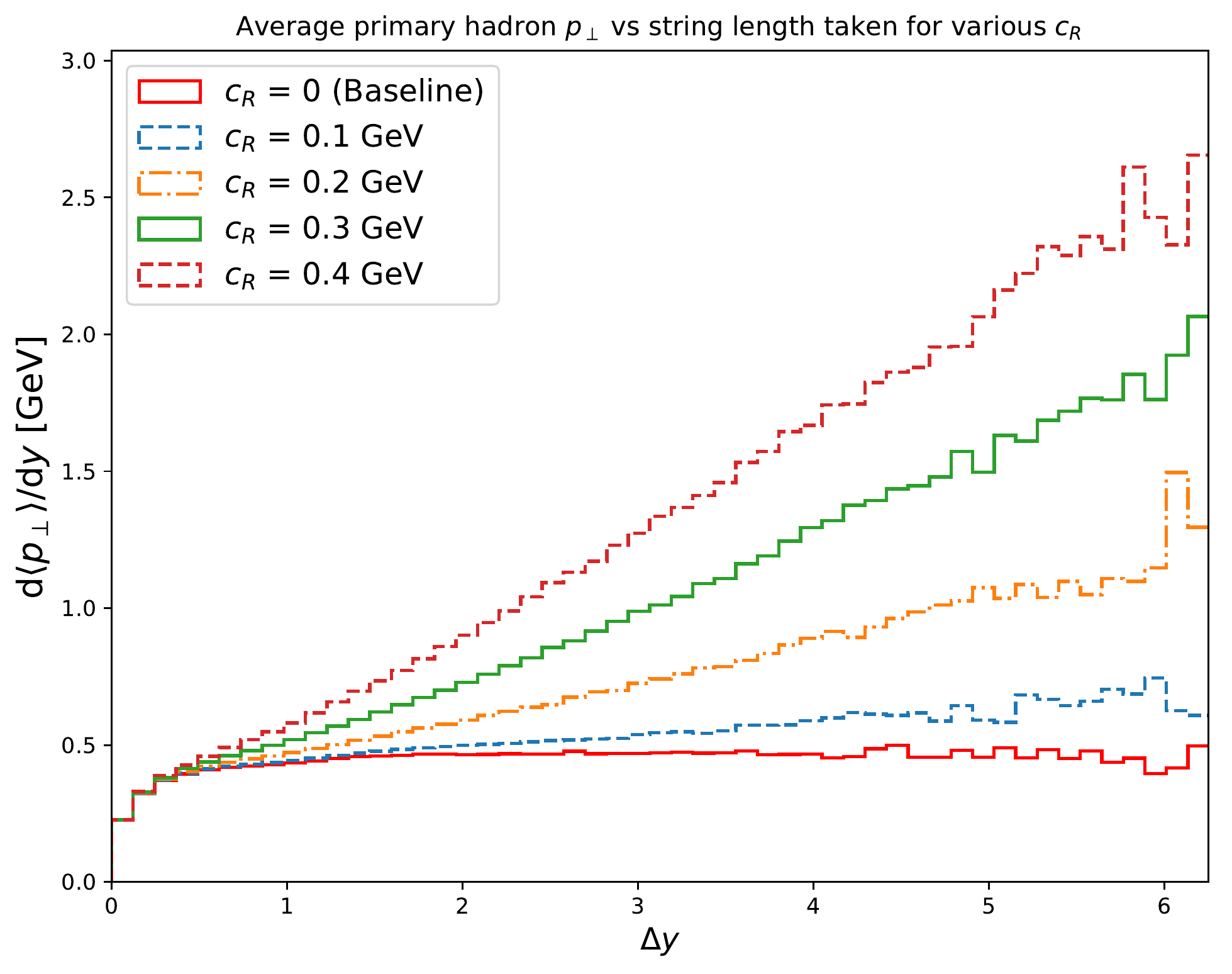}
\caption{ Distribution of average primary-hadron \pT as a function
of  $\Delta y_h$. {\sl Left:} comparison of baseline Lund model (red solid line) to our model for $c_R = 0.2$~GeV (blue solid line), our model with only the repulsion \pT component (blue dot-dashed line) and the shoving model (black dashed line). The shoving model
exhibits a lower average \pT since the soft gluons it adds make the strings longer causing the multiplicity of produced hadrons to increase faster than the total \pT. {\sl Right:} the effect of varying the repulsion strength $c_R$. }
\label{fig:proof}
\end{figure*}

In the rest of this section, we study the consequences of our model for an explicit example configuration defined by: 
\begin{equation}
    \begin{split}
        p_{+1} = p_{+2} &= 400\left(1,0,\vec{0}_{\perp} \right)\,\mathrm{GeV}~,\\
        p_{-1} = p_{-2} &= 400\left(0,1,\vec{0}_{\perp} \right)\,\mathrm{GeV}~.\\
    \end{split}
    \label{eq:symmetricConfig}
\end{equation}
To highlight the effects of the fragmentation repulsion, we have chosen endpoint energies of 200~GeV (corresponding to rather long strings), and, at this stage, consider only primary hadrons (hadrons that are
produced directly from the fragmenting string).  The smearing caused by decays of (short-lived) primary hadrons into secondaries will be discussed in Sec.~\ref{sec:finalState}.

The left pane of Fig.~\ref{fig:proof}
shows the average \pT 
of primary hadrons as a function of $\Delta y_h$, as defined by Eq.~(\ref{eq:simpleRap}).
The red dashed histogram shows the results of using the ordinary Lund model, which --- since the 
Gaussian transverse momentum generation in the baseline Lund model is independent of the rapidity span
--- is a flat distribution modulo endpoint effects, 
The two blue histograms illustrate the effects of our compression and fragmentation repulsion model, for a representative value of $c_R = 0.2$~GeV. 
The dot-dashed histogram shows the repulsion component by itself (obtained by turning off the Gaussian fragmentation \pT component via {\tt StringPT:sigma = 0}). The solid blue histogram shows the combination of the fragmentation and repulsion \pT components, for the same reference value of $c_R$. For small $\Delta y$, this mimics the baseline string model, while for large $\Delta y$, the repulsion \pT takes over as 
the dominant source of transverse momentum.

We also include a comparison to the shoving model as implemented in Pythia 8.2
\cite{Bierlich:2016vgw,Bierlich:2017vhg}. For the shoving parameters used in our study 
(see App. \ref{app:shoving} for details), the average transverse momentum per unit rapidity span taken actually decreases relative to the baseline (solid red) model.
We interpret this as a result of the physical mechanism by which the shoving model 
pushes the two strings apart, which is implemented as a number of very
soft transverse gluon excitations. While this does increase the total \pT, it also 
increases the total string length. The latter in turn increases the hadron multiplicity, with the result that 
the average \pT  per hadron can decrease. In our model, by contrast, the compression step ensures that the total multiplicity decreases; the repulsion step then adds \pT, implying that both the total and the average \pT per hadron must increase. 
 
The results of varying $c_R$ from 0~GeV (equivalent to the no-repulsion baseline case) to 0.4~GeV per 
unit of rapidity overlap 
are shown in the right panel of Fig.~\ref{fig:proof}. 
As $c_R$ increases, the slope of the average
hadron \pT increases with the rapidity span of the string taken, as expected
from the ansatz in Eq.~(\ref{eq:pThadron}).

\begin{figure*}[tp]
\centering
\includegraphics[width=0.49\textwidth]{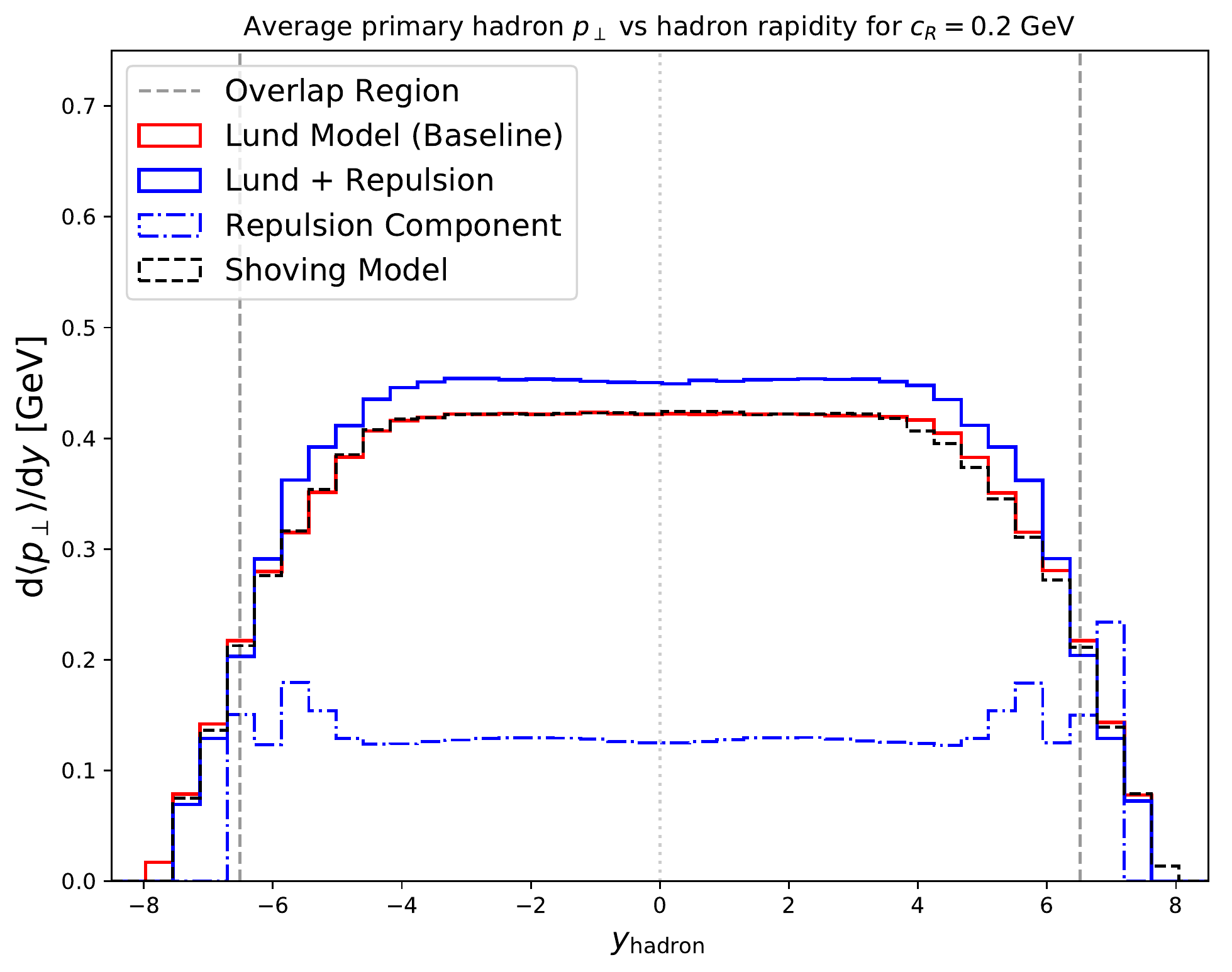}
\includegraphics[width=0.49\textwidth]{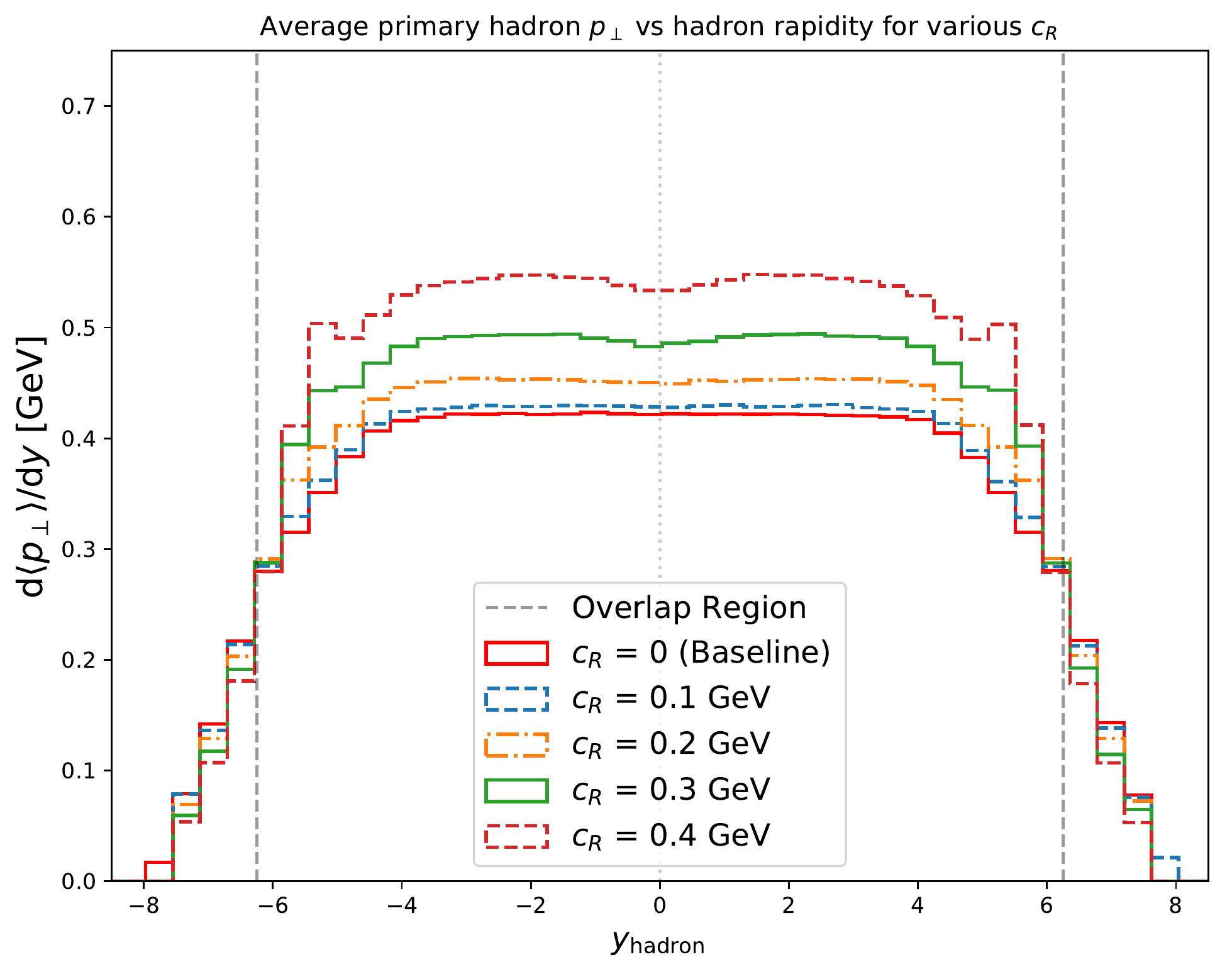}
\caption{Distributions for the average \pT of primary hadrons as a function of the
hadron's rapidity for the symmetric parallel strings configuration.
\textit{Left}: comparison of the baseline Lund model (red solid line),
with our fragmentation repulsion model (blue solid line),
which has a higher $\left< p_\perp\right>$ in the plateau region. 
The component which is due to the repulsion effect is illustrated by the 
blue dot-dashed line. Also shown is the result of using the 
shoving model (black dashed line)
\cite{Bierlich:2017vhg}, for the same string configuration. The shoving model
does not have significant deviation from the baseline Lund model for this observable (see
text).
\textit{Right}: the effect that varying the repulsion strength $c_R$.
}
\label{fig:plateau}
\end{figure*}

In Fig.~\ref{fig:plateau}, we show the same model examples but now as a function of the more directly observable rapidity of the hadrons, instead of the rapidity span they take.  
For the normal Lund string model, this produces a variant of the famous rapidity plateau (red solid line). For the parameters we studied, the shoving model (dashed black line) does not change the average $p_\perp$ appreciably (while the average multiplicity of the event is increased~\cite{Bierlich:2017vhg}). 
In contrast, for our reference value of $c=0.2$~GeV, our repulsion model (blue solid line, with the repulsion component illustrated by the blue dashed line) 
does increase the average primary hadron \pT. The net increase is less than linear since the ordinary (Gaussian) fragmentation \pT is oriented randomly with respect to the repulsion \pT, and the two components add vectorially. 

As in the previous figure, the right panel of Fig.~\ref{fig:plateau} illustrates the effect of varying $c_R$ in the range 0 to 0.4 GeV per unit of rapidity overlap. For larger values of $c_R$, the rapidity plateau begins to lose some of its flat structure, particularly in the middle of the string, near $y_{\rm hadron} = 0$. To fix the flatness, one may adjust the stopping mass parameter $W^2_{\rm stop}$ in Pythia's implementation of the string model, though this is outside the scope of this work.

\section{General Parallel Two-String Configuration}
\label{sec:generalStrings}
We now extend the considerations in Sec.~\ref{sec:simplestStrings} to a more general
configuration, by letting the strings have an arbitrary parallel configuration. Without loss of generality, we assume that the two strings do still overlap, either partially, or one string's rapidity span is fully contained inside the rapidity span of the other. Relabeling as needed, we require in the former case that the left-moving ($W_- $) end of string 1 is contained within the rapidity span of string 2,  and the right-moving ($W_+ $) end of string 2 is contained within the rapidity span of string 1. 

In the context of the momentum-space representation of the Lund model that our repulsion framework is based on, the full space-time evolution of a string is determined solely by the starting values of the 4-momenta of its endpoints. By initially reducing these momenta, the ``compression'' step of our model expresses the physical expectation that, as two nearby strings expand simultaneously and repel each other, it will not be possible to convert \emph{all} of the kinetic energy of their endpoints into potential energy stored in the corresponding strings; instead, some fraction of the original kinetic energy is ``held in reserve'', to be converted into transverse momentum during the fragmentation process. When we now turn to consider asymmetric configurations, we must answer not only how much of the total kinetic energy must be held in reserve in this way, but also which fraction of it to take from each of the reservoirs represented by the two endpoints.  

\begin{figure}[tp]
\centering
\includegraphics[width=0.45\textwidth]{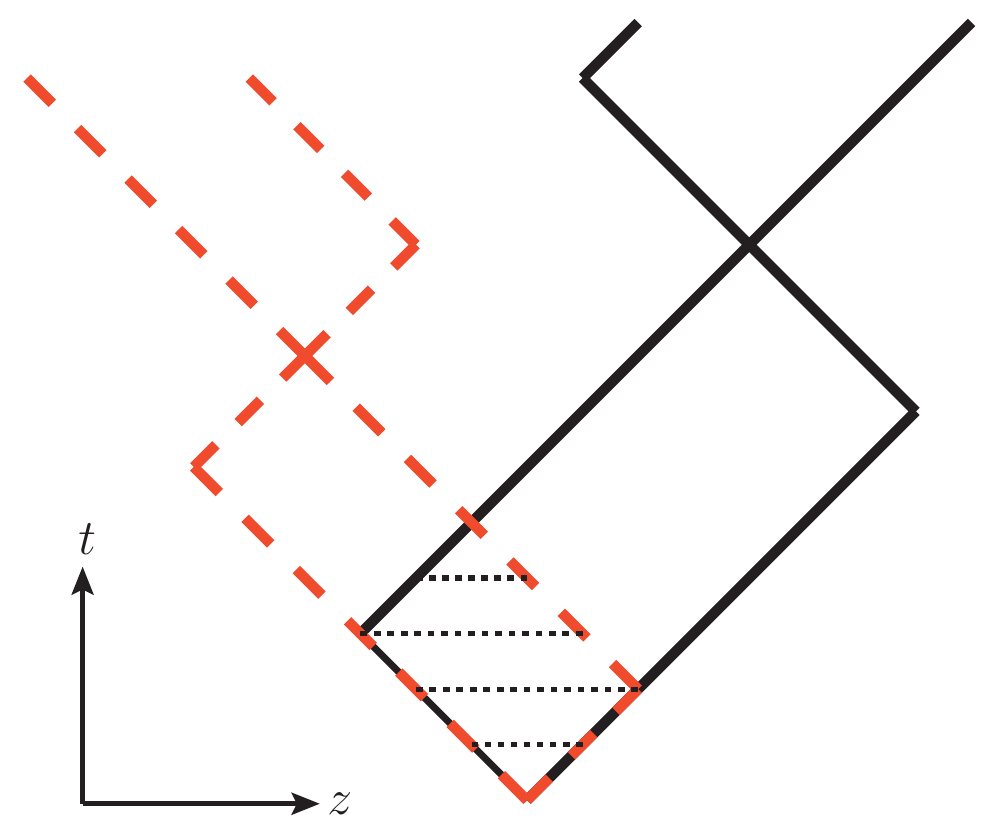}

\caption{ Schematic (1+1)-D spacetime diagram of the general
parallel two-string configuration,
where the two strings have a region of overlap 
(dotted parallel lines). The two endpoints 
in the region of overlap will be subjected to more
compression and repulsion.
}
\label{fig:generalSpacetime}
\end{figure}

In our fragmentation repulsion model, we will use the ansatz that endpoints ``inside'' a region of overlap should undergo more compression than ones ``outside'', since the corresponding string regions experience more of the accumulated interaction. In Fig.~\ref{fig:generalSpacetime}, we show a (1+1)-D diagram of a general string configuration, with an overlapping region centred around a slightly negative rapidity (in the given frame). The right-moving endpoint of the dashed-orange string piece overlaps with the solid-black string system during the entire time over which its original kinetic energy is converted to potential energy. By contrast, the left-moving endpoint of the same  dashed-orange string piece only overlaps with the black system during half of the time that it takes to convert all of its kinetic energy to potential energy. In this sense, the right-moving endpoint can be considered to be ``inside'' the region of overlap while the left-moving one ultimately travels ``outside'' of that region. Alternatively, the portion of the black-solid string system that is represented by its left-moving endpoint has a bigger fraction of total overlapping area than the portion that is represented by its right-moving endpoint. 

\subsection{String Compression}
In the general case that the strings are not symmetric in the longitudinal direction, one must make a choice whether to allow them to exchange $p_L$ or not. For simplicity and since we wish to focus on the 
transverse repulsion effects here, we choose to ignore 
the possibility of $p_L$ exchange in this first 
version version of our model. Thus, the only change with respect to the symmetric case is that the rescaling factors for each of the four endpoint momenta will no longer be equal. 

Regardless of longitudinal recoil, the compression factors for each string system $i\in [1,2]$ must satisfy: 
\begin{equation}
\begin{split}
    f_{+1}f_{-1} &= f^2_1 = 1 - \frac{p_{\perp,R}^2}{W_1^2}~, \\
    f_{+2}f_{-2} &= f^2_2 = 1 - \frac{p_{\perp,R}^2}{W_2^2}~, \\
\end{split}
\label{eq:constraints}~,
\end{equation}
where $p_{\perp, R}$ is the total $p_\perp \propto \Delta y_\mathrm{ov}$ from repulsion to be assigned (equally and oppositely) to the two systems, see eq.~(\ref{eq:pTrepel}), 
and longitudinal momentum conservation, $\Delta p_{L,1} = -\Delta p_{L,2}$, implies:
\begin{equation}
\begin{split}
    \left(1-f_{+1}\right)&W_{+1} - \left(1-f_{-1}\right)W_{-1}  \\
    &=\left(1-f_{-2}\right)W_{-2} - \left(1-f_{+2}\right)W_{+2}~.
\end{split}\label{eq:pLcons}
\end{equation}
This gives three constraints for four unknowns. Imposing the further condition of no longitudinal momentum exchange, $\Delta p_{L,1}=\Delta p_{L,2}=0$,   eq.~(\ref{eq:pLcons}) separates into:
\begin{equation}
    \begin{split}
        \left(1-f_{+1}\right)W_{+1} - \left(1-f_{-1}\right)W_{-1} = 0 ,\\
        \left(1-f_{-2}\right)W_{-2} - \left(1-f_{+2}\right)W_{+2} = 0.
    \end{split}
    \label{eq:internal}
\end{equation}
The problem can then be solved with a unique set of solutions for each compression
factor $f_{\pm i}$. Inserting the first two constraints Eq.~(\ref{eq:constraints}) into
Eq.~(\ref{eq:internal}), we obtain a quadratic equation
for $f_{-i}$:
\begin{equation}
    W_{-i}f_{-i}^2 + \left(W_{+i} - W_{-i}\right)f_{-i} - f_i^2W_{+i} = 0 .
\end{equation}
Since the compression factors must be positive, there is only one solution
to this equation:
\begin{equation}
    f_{-i} = \frac{\left(W_{-i} - W_{+i}\right) + \sqrt{\left(W_{-i} - W_{+i}\right)^2 + 4W_i^2f_i^2}}{2W_{-i}} ,
    \label{eq:internalSolution}
\end{equation}
or equivalently using the longitudinal momentum component $W_{Li} = (W_{+i} - W_{-i})/2$,
\begin{equation}
    \begin{split}
   W'_{-i} = f_{-i}W_{-i} =  \sqrt{W_{Li}^2 + W_i^2f_i^2} - W_{Li}~,\\
   W'_{+i} = f_{+i}W_{+i} =  \sqrt{W_{Li}^2 + W_i^2f_i^2} + W_{Li}~.
\end{split}
    \label{eq:internalSolutionL}
\end{equation}
In the limit of $W_{+i} = W_{-i}$, i.e.\ $W_{Li}=0$, we reproduce the 
symmetric case for the given string $i$, i.e. 
$f_{\pm i} = \sqrt{f^2_i}$.
By construction, longitudinal momentum is conserved, $W'_{+i} - W'_{-i} = W_{+i} - W_{-i}$. 
However, energy is not:
\begin{equation}
 E'_i  = \frac{W'_{+i} + W'_{-i}}{2}
  = E_i\sqrt{1 - \frac{p_{\perp, R}^2}{E_i^2}}~. 
\end{equation} 
When we perform the fragmentation repulsion, we regain the ``lost'' energy by
giving the primary hadrons the repulsion \pT and putting them on-shell again,
with the string remnant absorbing the remaining energy. Thus, we conserve
energy and momentum after compression \textit{and} fragmentation of the strings.

It should be mentioned that our choice of no $p_L$ exchange does introduce a dependence on the frame in which the system is considered. 
This is due to the fact that while the lightcone
momenta $W_{\pm}$ follow a simple rescaling under longitudinal boosts, 
the compression
factors $f_{\pm i}$ depend non-linearly on $W_{\pm i}$ as seen in Eq.~(\ref{eq:internalSolution}), complicating their transformations under such boosts.
Specifically, compressing the strings then boosting the entire system results in a (marginally) different momentum topology than boosting the strings with the same boost factor and then compressing them. In this work unless otherwise stated, we compute compression factors in the overall CM frame of the two-string system. (A possible alternative, not pursued here, would be to  boost the system  longitudinally such that the centre of the overlap region is at $y= 0$.)

\subsection{Repulsion}
\label{sec:generalRepulsion}
\begin{figure*}[tp]
\centering
\includegraphics[width=0.99\textwidth]{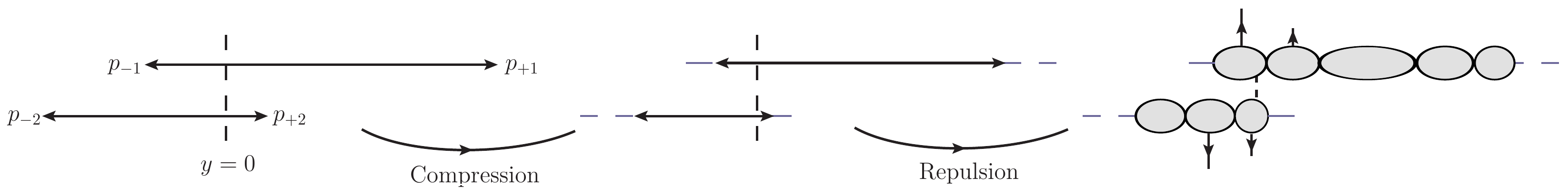}

\caption{ Schematic diagram of the general two-string configuration
and how we perform the string compression using 
Eq.~(\ref{eq:internalSolution}), and then the fragmentation
repulsion. Only primary hadrons in the region of overlap will receive \pT proportional
to the string length taken. We have ignored the Gaussian transverse momentum
generation in the Lund string model for the purposes of this figure.
}
\label{fig:generalStringsRepulsion}
\end{figure*}
The repulsion effect we seek to model is local;  additional \pT should be imparted to hadrons formed within regions of string overlap, and not to those outside. Fragmenting the (compressed) string from the outside in as usual, and using Eq.~(\ref{eq:simpleRap}) to compute rapidity spans, we distinguish three cases for each produced hadron:  
\begin{enumerate}
    \item The span is completely outside the overlap region;
    \item The span is completely inside the overlap region;
    \item The span straddles the boundary of the overlap region.
\end{enumerate}
In the first case, the hadron receives no repulsion \pT, while in the second, it is computed according to Eq.~(\ref{eq:pThadron}) and assigned repulsion
\pT following the
same procedures as described in Sec.~\ref{sec:simplestStrings}. 
In the last case,
only the portion of the rapidity
span inside the overlap region contributes to Eq.~(\ref{eq:pThadron}).

To illustrate the repulsion effect we consider a two-string scenario defined by the following endpoints (using the same lightcone notation as previously),
\begin{equation}
    \begin{split}
        p_{+1} &= 1200\left(1,0,\vec{0}_{\perp} \right)\,\mathrm{GeV} ,\\
        p_{-1} &= ~300\left(0,1,\vec{0}_{\perp} \right)\,\mathrm{GeV} ,\\
        p_{+2} &= ~100\left(1,0,\vec{0}_{\perp} \right)\,\mathrm{GeV} ,\\
        p_{-2} &= 1000\left(0,1,\vec{0}_{\perp} \right)\,\mathrm{GeV} ,\\
     \end{split}
    \label{eq:generalConfig}
\end{equation}
This configuration is then boosted back to the overall CM frame. 
An illustration of the compression and repulsion steps for this type of configuration is given in Fig.~\ref{fig:generalStringsRepulsion}. 

\begin{figure*}[t]
\centering
\includegraphics[width=0.49\textwidth]{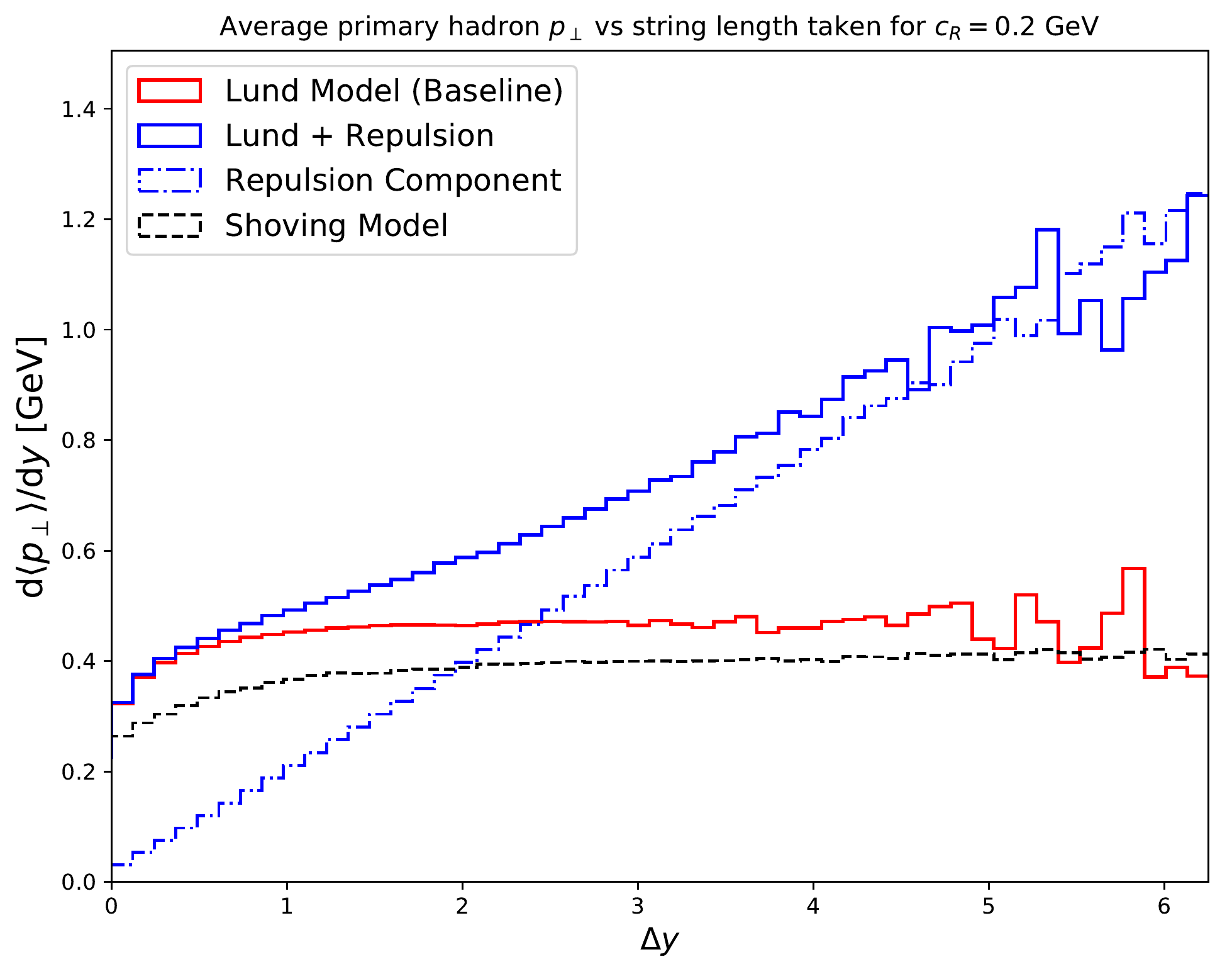}
\includegraphics[width=0.49\textwidth]{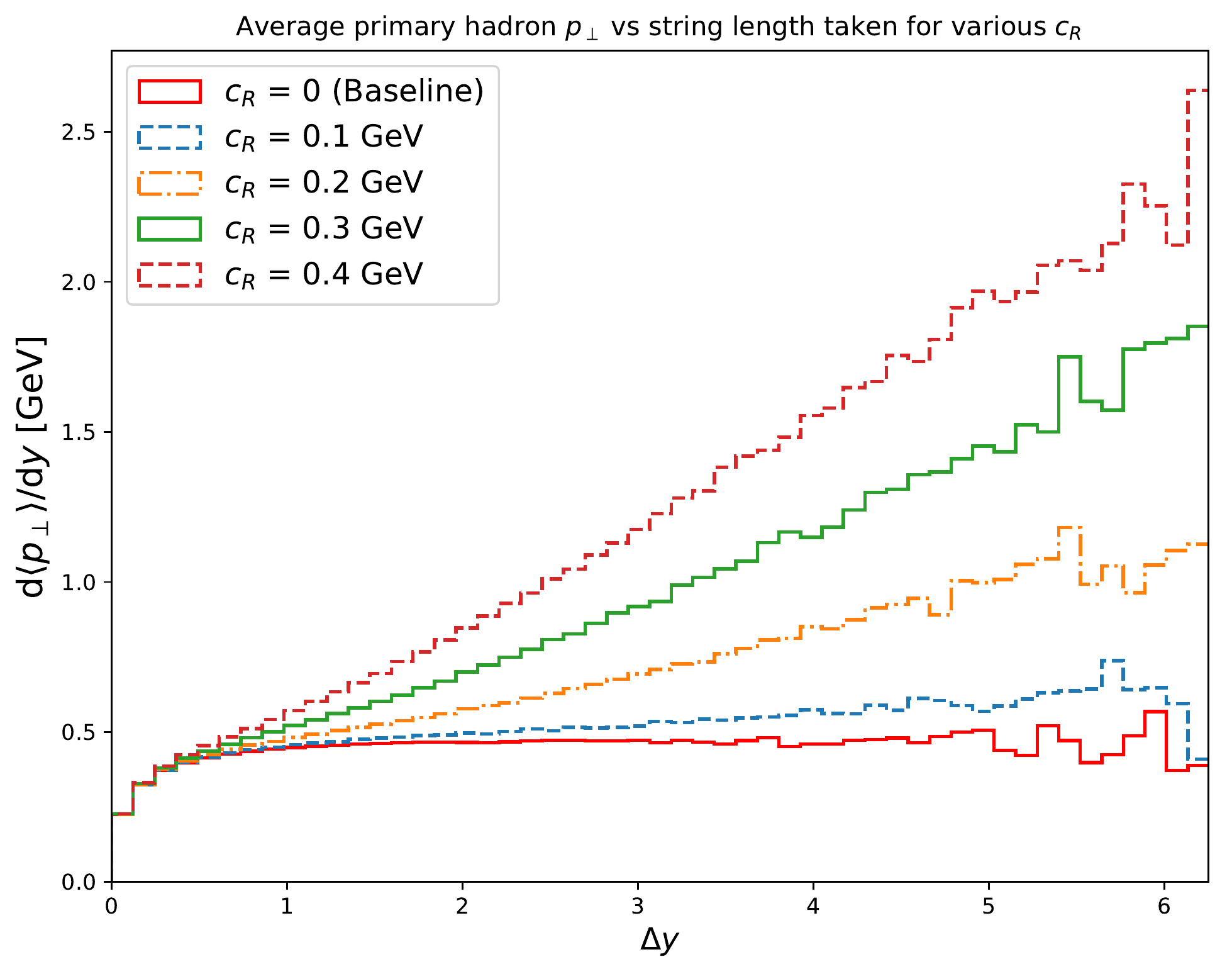}
\caption{ Distribution of average hadron \pT for primary hadrons as a function
of the rapidity span that they take, for the
asymmetric two-string configuration discussed in the text. \textit{Left}: the baseline 
Lund model (red solid) compared to our model of fragmentation repulsion (blue solid line).
For the latter, the blue dot-dashed histogram illustrates the component which is due to repulsion. 
Also shown is the result of using the shoving model, which, like the baseline Lund model, is agnostic to the amount of string length taken.
\textit{Right}: the effect of varying $c_R$ in Eq.~(\ref{eq:pThadron}).
}
\label{fig:proofGeneral}
\end{figure*}
\subsection{Results}

In Fig.~\ref{fig:proofGeneral}, we show the average
primary hadron \pT distribution as a function of the 
string rapidity span taken by the hadron. 

In the left panel of Fig.~\ref{fig:proofGeneral}, 
the red histogram is the
ordinary Lund model, which is agnostic to the the rapidity span
taken by a hadron. 
The blue histograms are the result of our implemented 
model for $c_R = 0.2$~GeV, for both the repulsion component (dot-dashed), which
matches the ansatz in Eq.~\eqref{eq:pThadron}, and
the full fragmentation (solid), which matches the baseline Lund model and
the repulsion component in the limits of small and large $\Delta y$ respectively.
Lastly, we have also included the results of using the shoving model
for this configuration. These results are largely similar to the results
for the symmetric, parallel configuration in Sec.~\ref{sec:simpleResults}.

The right panel of Fig.~\ref{fig:proofGeneral} highlights the
effects of varying $c_R$ on the average primary hadron 
\pT distribution
as a function of the string's rapidity span taken by the hadron
for the full fragmentation
repulsion. For $c_R = 0$~GeV, we reproduce
the ordinary Lund model. As the repulsion factor increases,
the slope of the average \pT increases, since the two are proportional
via Eq.~(\ref{eq:pThadron}).

\begin{figure}[tp]
\centering
\includegraphics[width=0.47\textwidth]{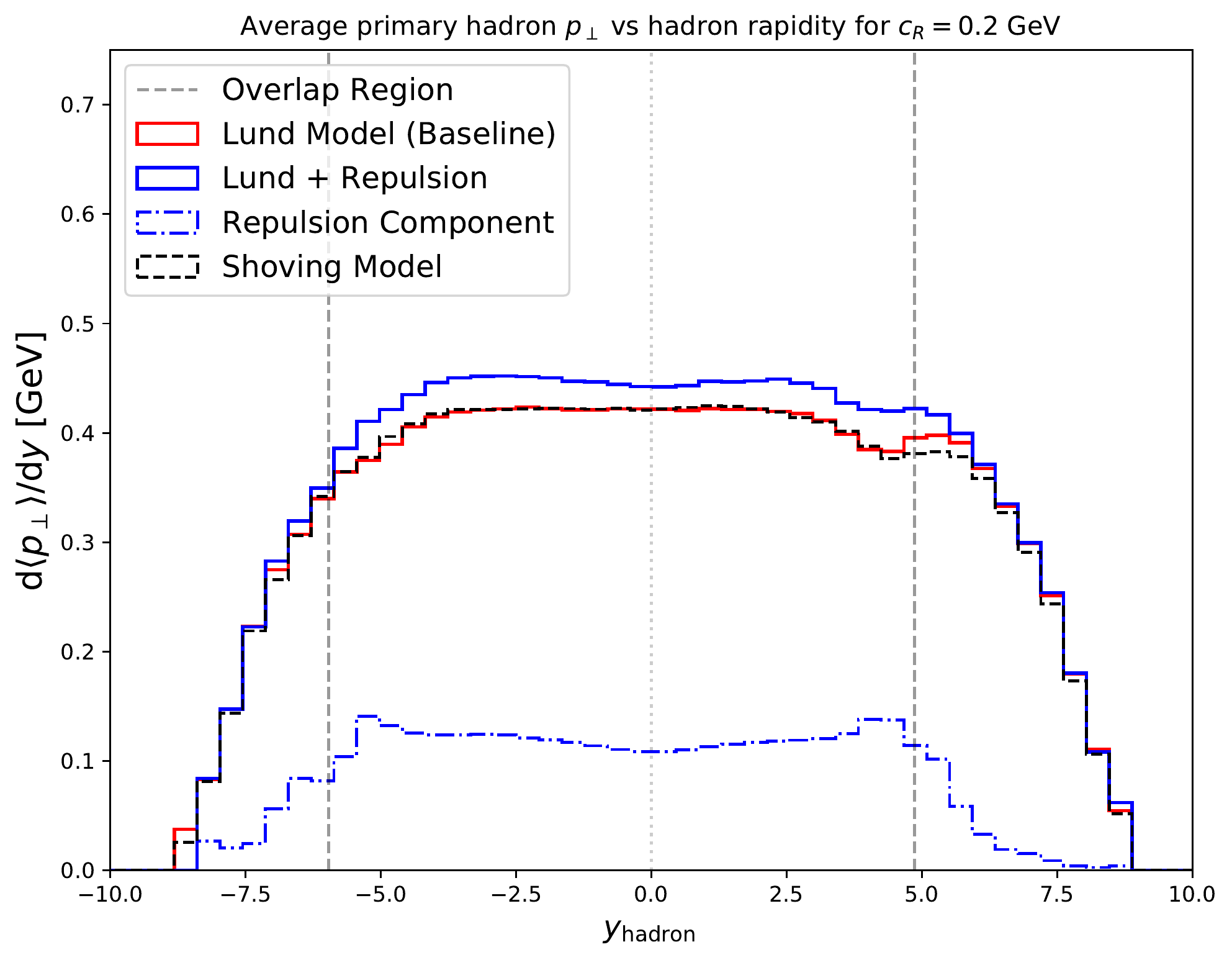}
\caption{Distribution of average hadron \pT for primary hadrons as a function
of $y_\mathrm{hadron}$, for the asymmetric two-string example described in the text. 
The repulsion component of our fragmentation repulsion increases the $\left<p_\perp\right>$ in the region of overlap (indicated by the grey dashed lines, using $m_0 = 0.5$ GeV in the rapidity calculation).
}
\label{fig:avgpTGeneral}
\end{figure}

In Fig.~\ref{fig:avgpTGeneral}, we present the average primary 
hadron \pT distribution as a function of the hadron's rapidity 
(as measured in the overall CM frame of the two strings).
Since the configuration is asymmetric with
respect to the endpoints of the two strings, the resultant
compression and fragmentation repulsion will also reflect this
asymmetry. The red histogram is the ordinary Lund string model, and
again we reproduce the rapidity plateau, with a small asymmetry
due to the configuration of strings.
The blue histograms are our fragmentation repulsion for
$c_R = 0.2$~GeV where we have shown only the repulsion component (dot-dashed),
and the full fragmentation repulsion (solid). 
We have also included the results of the shoving model (black dashed).

Fig.~\ref{fig:avgpTGeneral} also showcases the considerations from
Sec.~\ref{sec:generalRepulsion}. In comparison to Fig.~\ref{fig:plateau} 
where the repulsion component has a sharp cut-off
at the edges of the rapidity overlap region, in the general case we
have longer tails that extend beyond the overlap region due to hadrons
taking rapidity spans that are only partially in the
overlap region.

Comparing Figs.~\ref{fig:plateau} 
and \ref{fig:avgpTGeneral}, we see the same structures for each respective
model. Our fragmentation repulsion exhibits an increased average \pT for hadrons inside the 
rapidity overlap region, while hadrons outside that region have
a diminished \pT contribution from the repulsion. As in the previous section, we see that the shoving model considered in this study does not change the distribution, apart from minor deviations near the endpoints.

\section{Two-String Systems with Relative Rotations and Boosts}
\label{sec:boosted}
\begin{figure*}[tp]
~~~~~\includegraphics[width=0.4\textwidth]{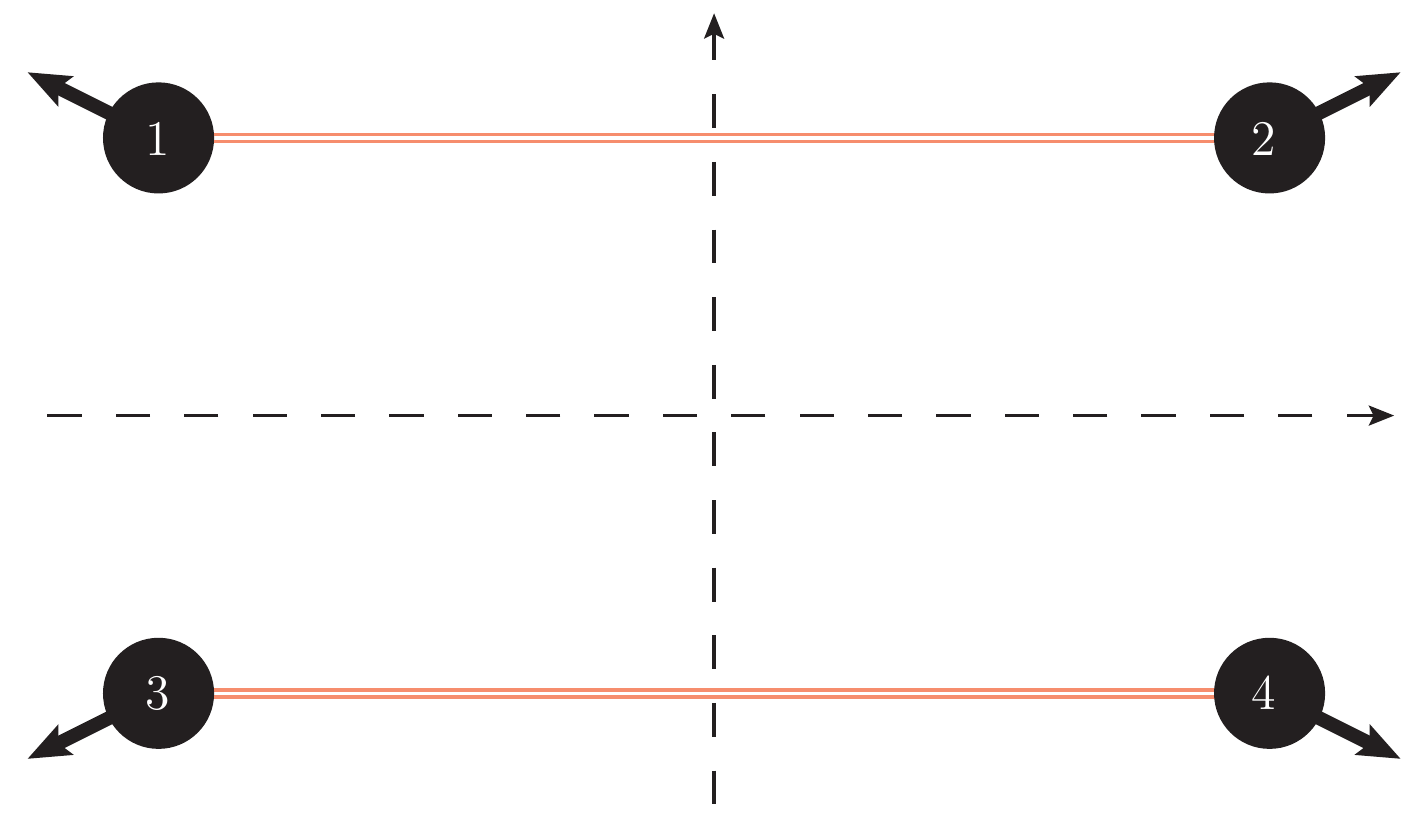}\hfill%
\includegraphics[width=0.4\textwidth]{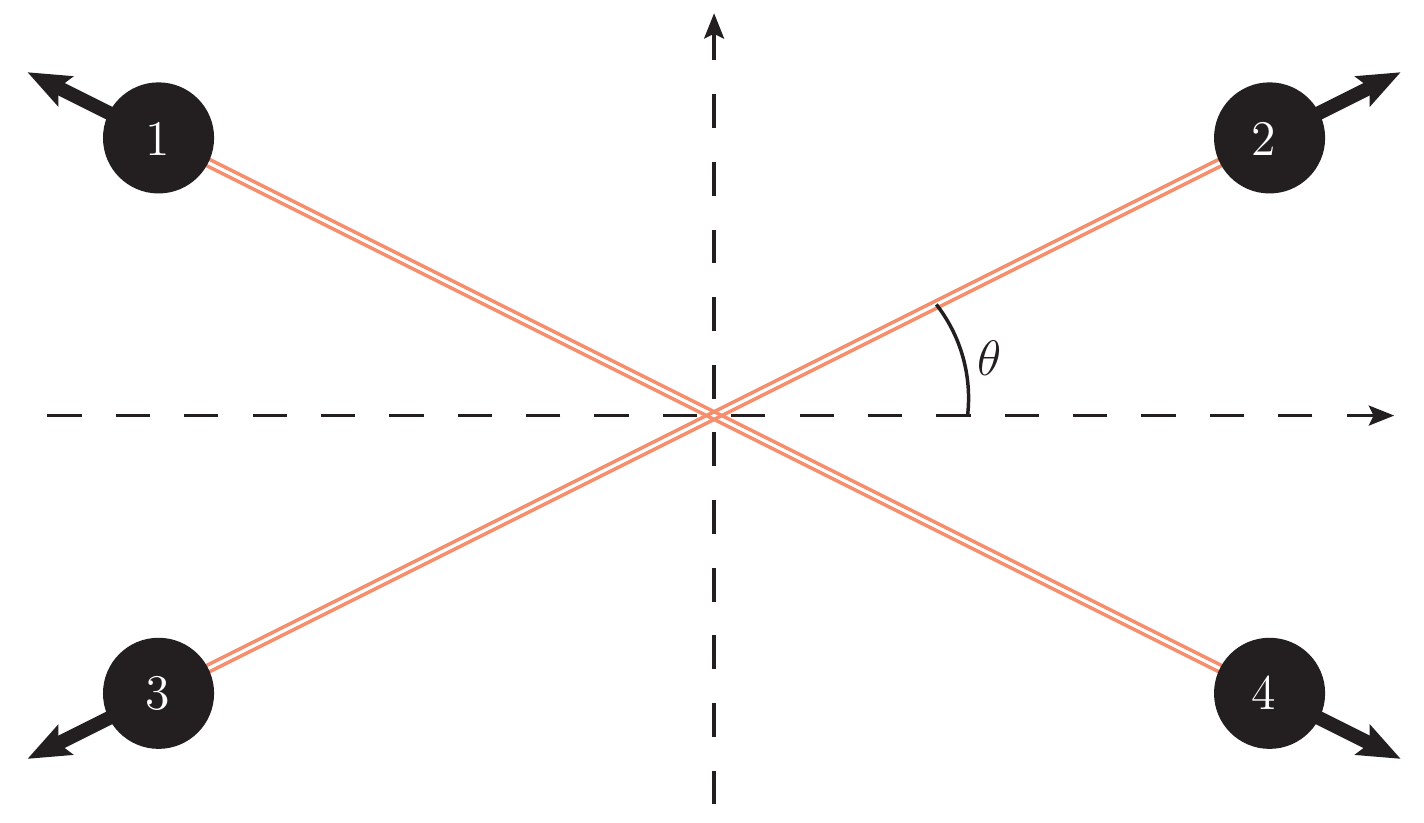}~~~~~
\caption{Schematic diagram of two string systems defined by the 
 endpoint momenta given in Eq.~(\ref{eq:probo}), 
corresponding to (left) a relative boost and (right) a relative rotation.}
\label{fig:robo}
\end{figure*}

We now consider string systems with endpoints that have non-vanishing 
transverse momenta. The examples we consider in this section will still be defined 
so that the $z$ axis remains a sensible choice of common rapidity axis. Specifically, 
we will consider systems like those illustrated in Fig.~\ref{fig:robo}, 
with endpoint momenta (in conventional 4-momentum notation):
\begin{equation}\begin{array}{rcrrrr}
p_1 & = & E(~1,&\sin\theta,&~0,&-\cos\theta)~,\\
p_2 & = & E(~1,&\sin\theta,&~0,& \cos\theta)~,\\
p_3 & = & E(~1,&-\sin\theta,&~0,&-\cos\theta)~,\\
p_4 & = & E(~1,&-\sin\theta,&~0,& \cos\theta)~,
\end{array}\label{eq:probo}
\end{equation}
so that the string systems defined by the (1,2) and (3,4) pairings are still parallel but each are transversely boosted relative to the overall CM, by $\beta = \pm \sin\theta$, while the systems defined by the pairings (1,4) and (2,3) are at rest relative to the overall CM but are rotated with respect to each other, with a relative opening angle of $2\theta$. In all cases, the CM energy is $E_\mathrm{CM} = 4E$. For definiteness we take $\sin\theta= 0.1$ in the examples below unless otherwise stated. 

\subsection{Symmetric configuration with relative boost}
\label{sec:symmetricBoosted}
Taking the simplest symmetric two-string configuration, we ask
what happens in the situation depicted in the left-hand panel of Fig.~\ref{fig:robo} in which 
both strings have some (equal and opposite) transverse momentum before compression. 

Using the same arguments as above, we wish to convert a fraction of the original longitudinal momenta of the endpoints (defined along the common rapidity axis, here the $z$ axis) 
into transverse momentum instead of into potential energy of the string(s). 

As before, the total amount of repulsion $p_\perp$ is determined from the effective rapidity overlap, which we compute from the longitudinal momentum components (along the chosen common axis) of the endpoints, 
\begin{equation}
\Delta y_{\rm ov} = \mathrm{min}(y_{1+},y_{2+}) - 
\mathrm{max}(y_{1-},y_{2-})~,\\
\label{eq:overlap}
\end{equation}
where $y_{i+}$ and $y_{i-}$ refer 
to the rapidities of the right- and left-moving 
endpoints of string $i$ respectively and we regulate the rapidity values of massless endpoints in the $p_\perp \to 0$ limit by imposing $m\ge m_0$ in the denominator of our rapidity definition: 
\begin{equation}
    y = \ln \frac{E+p_L}{\sqrt{m^2 + p_\perp^2}} ~.
\end{equation}
Using lightcone coordinates as before, the longitudinal component of a general 
string-end momentum is $p_L = (W_+ - W_-)/2$, and the energy is $E = (W_+ + W_-)/2$. 

The string-ends will be rescaled
in a similar manner to the parallel strings in Sec.~\ref{sec:simplestStrings}. Since the rescaling is done on the full 4-vectors, the string endpoints will lose some $p_{\perp}$. 
We use the ansatz of giving this
extra transverse momentum reservoir, denoted $p_{\perp,{\rm res}}$ to 
the fragmenting hadrons as a fraction of the rapidity span
they take from the string:
\begin{equation}
    p_{\perp,{\rm h}} = \left(c_R + \frac{p_{\perp,{\rm res}}}{\Delta y_{\rm ov}}\right) \Delta y_h ,
    \label{eq:pThadronTrans}
\end{equation}
where $\Delta y_{\rm ov}$ is string-string overlap defined via Eq.~(\ref{eq:overlap}, and $\Delta y_h$ is the amount of rapidity
span taken by the hadron inside of the overlap region, as discussed
in Sec.~\ref{sec:generalStrings}. (Alternatively, and probably more correctly, one could distribute $p_{\perp,{\rm res}}$ among all the hadrons, not just those in the overlap region; or boosting the compressed string transversely so that it regains its original total $p_\perp$; but since since $p_{\perp,{\rm res}}$ is typically very small it is a minor effect.)

As in the previous section we assume no longitudinal momentum exchange, $\Delta p_{\rm L} = 0$. Writing the total longitudinal momentum of string $i \in [1,2]$ as  
\begin{equation}
     p_{L,i} = p_{L,+i} + p_{L,-i} ~,
\end{equation}
with $p_{L,\pm i}$ the longitudinal momentum of the respective endpoints, we can generalize  Eq.~(\ref{eq:internalSolution}) to:
\begin{equation}
    f_{-i} = \frac{ p_{L,i} + \sqrt{p^2_{L,i} - 4p_{L,-i}p_{L,+i}f_i^2}}{2p_{L,-i}} .
    \label{eq:transInternalSolution}
\end{equation}
In the limit of the string ends carrying \pT $\to 0$, Eq.~(\ref{eq:transInternalSolution}) exactly reproduces Eq.~(\ref{eq:internalSolution}).

The amount of repulsion $\perp$ given to each hadron during the fragmentation process should be proportional to the (overlapping portion of the) rapidity span it takes. 
The definition, Eq.~(\ref{eq:hadronspan}), is given in terms of the quantities used to characterize the fragmentation of each string in its own CM frame, along the axis defined by its endpoints in that frame, 
whereas we here want to along the chosen common axis in the string-string CM frame. 
As a very simple way to ``project'' the rapidity span, we use 
\begin{equation}
    \Delta y_{\rm eff} = \frac{\Delta y_\mathrm{string}}{\Delta y_\mathrm{string}^*} \Delta y^*_{\rm taken}~,
    \label{eq:yEffective}
\end{equation}
where the $\Delta y_\mathrm{string}$ is the rapidity span of the given string evaluated along the common axis defined in the string-string frame and $\Delta y^*_\mathrm{string} = \ln\left(W^2/m^2_0\right)$ is the (larger) span evaluated in the string's own rest frame. $\Delta y^*_{\rm taken} = \ln\left(W^2/W'^2\right)$ is the rapidity
span of the hadron taken in the string's own rest frame.

The effective string length in Eq.~(\ref{eq:yEffective}) taken is invariant under longitudinal boosts, and
reproduces the parallel configuration in the limit where each string endpoint carries
vanishing $p_{\perp}$.
Eq.~(\ref{eq:yEffective}) also sums to give the correct rapidity span
along the $z$-axis, 
and is agnostic to the direction of the 
transverse momentum.

The last point to address is in which direction in azimuth to apply the repulsion. Considering the transverse plane only (in the string-string CM frame), the two systems will have some equal and opposite overall motion, which we denote by $\vec{p}_{\perp,\mathrm{rel}} = \vec{p}_{\perp 1} - \vec{p}_{\perp 2} = 2\vec{p}_{\perp 1}$. Assuming that, by the time strings are formed, the string systems are already separated a bit (on average) along this axis, it seems plausible to us to apply the repulsion $p_\perp$ along the same direction. To provide some variability and in order to have a well-defined repulsion axis also in the $p_{\perp,\mathrm{rel}}\ to  0$ limit, we add a random component as well:  
\begin{equation}
    \vec{n}_\mathrm{\perp 1} = N (\vec{p}_{\perp,\mathrm{rel}} + \rho \vec{n}_{\perp, \mathrm{ran}})
    \label{eq:nT}
\end{equation}
where $\vec{n}_{\perp,\mathrm{ran}}$ is a unit-vector in a randomly chosen azimuthal direction, the normalisation factor 
\begin{equation}
 N = \frac{1}{\sqrt{p_{\perp,\mathrm{rel}}^2 + \rho^2 + 2 \rho (\vec{p}_{\perp,\mathrm{rel}}\cdot \vec{n}_{\perp,\mathrm{ran}})}}
\end{equation}
ensures $|n_{\perp 1}| = 1$, and $\rho$ is a free parameter of order 1 GeV which governs the relative importance of the random component. The repulsion for string 1 is oriented \emph{with} $n_{\perp 1}$, and that for string 2 in the opposite direction.

The choice of direction can have a significant effect on two-particle azimuthal
correlations, as we will describe in Sec.~\ref{sec:cumulant}, but it does not
have a drastic effect at the level of the distributions for the average
hadron transverse momentum versus hadron rapidity and rapidity span taken.

\subsection{Results}
\label{sec:symmetricBoostedResult}
\begin{figure*}[tp]
\centering
\includegraphics[width=0.49\textwidth]{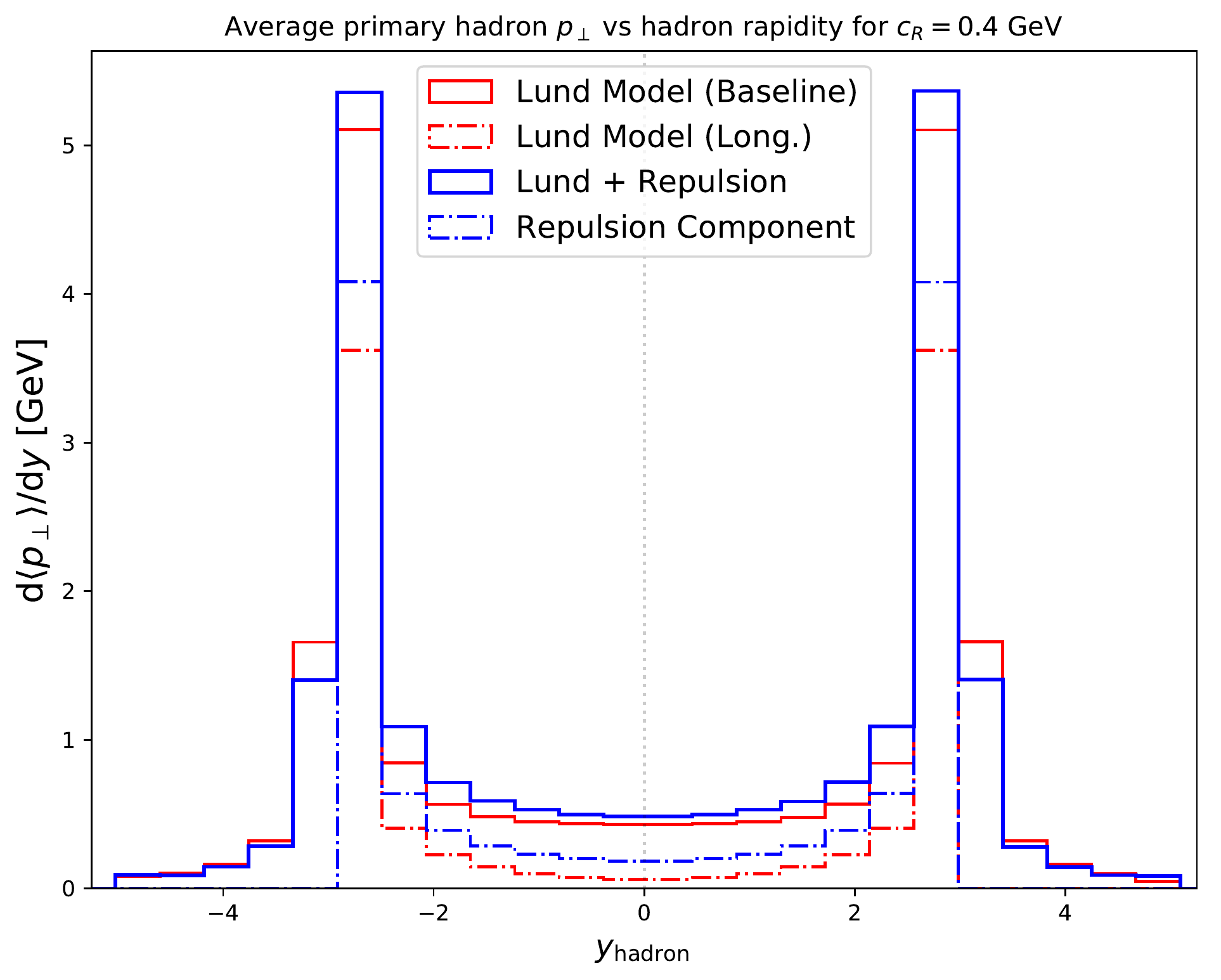}
\includegraphics[width=0.49\textwidth]{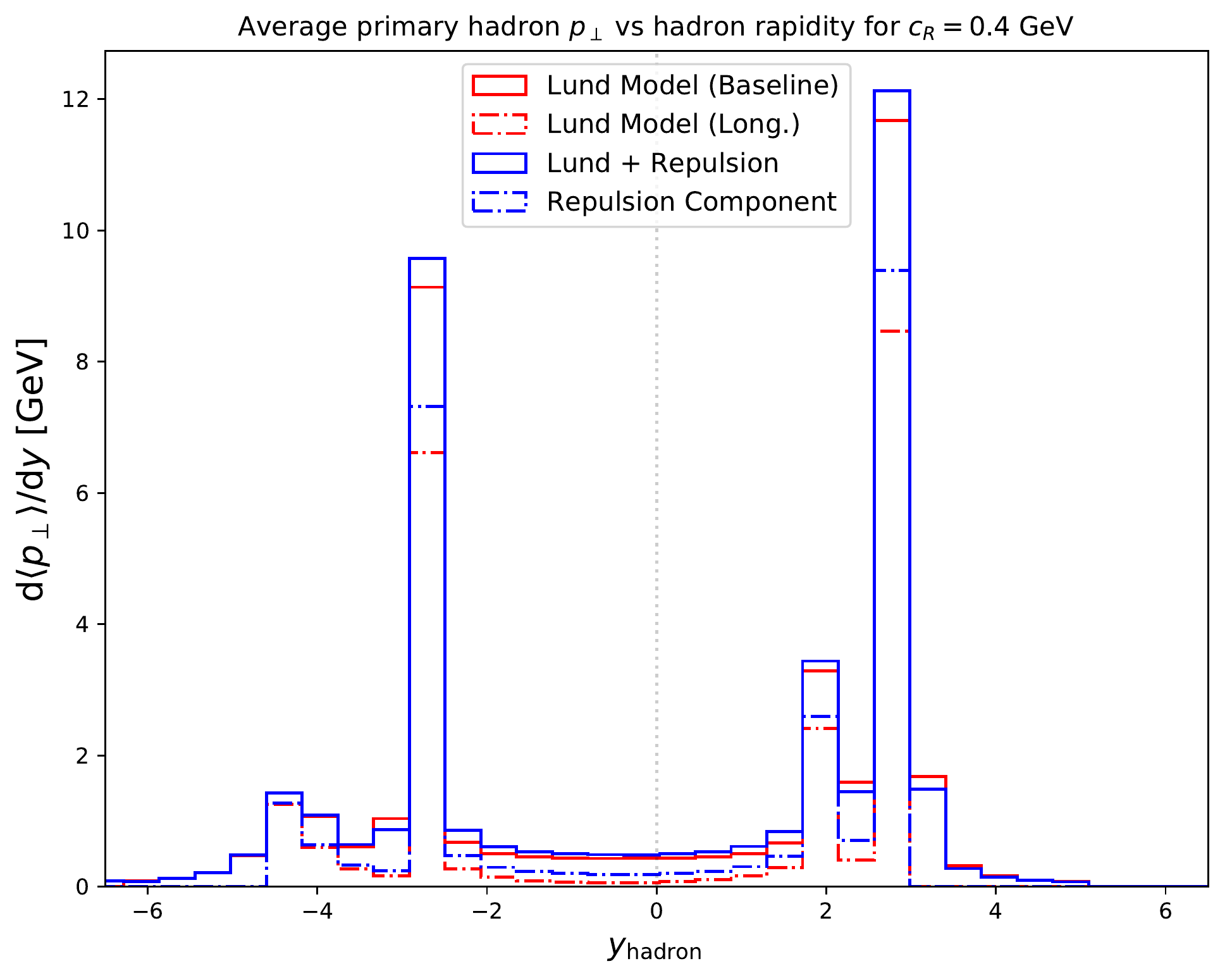}
\caption{Distribution of average hadron \pT of primary hadrons as a function
of the hadron's rapidity, for the symmetric configuration (left)
and the general configuration (right),
where the two strings have an equal and opposite boost
in the transverse direction. The latter configuration is boosted
back to the two-string rest frame before compression and fragmentation.
We have added the repulsion \pT in the same direction as the overall
motion of each string.
}
\label{fig:trans_average_pT}
\end{figure*}
We present the results in Fig.~\ref{fig:trans_average_pT}. Both panels
show the average primary hadron \pT as a function of $y_\mathrm{hadron}$ (defined along the common string-string axis, here the $z$ axis, in their overall CM frame). 
In these plots we have chosen to add the repulsion \pT
in the same direction as each string's overall transverse motion, and chosen
a larger value of $c_R = 0.4$~GeV compared to the parallel
configurations, to make the effects of the repulsion stand out a bit more clearly against the $p_\perp$ contributed already from the endpoints. 

In Fig.~\ref{fig:trans_average_pT}, our fragmentation repulsion
exhibits similar effects as those seen in Figs. \ref{fig:plateau} and
\ref{fig:avgpTGeneral}, though the enhancement of average primary hadron
\pT is less drastic than in the parallel configurations. The reason for
this is twofold: first, since the strings are no longer parallel,
the amount of rapidity overlap between the two strings is reduced, resulting in less total repulsion \pT. 
Second, the boosted endpoints show up as peaked structures around the endpoints' rapidity.

While the effects of our fragmentation repulsion are less distinctive
in Fig.~\ref{fig:trans_average_pT}, the framework will still have a
distinctive effect on the two-particle azimuthal correlations, 
as we will discuss in Sec.~\ref{sec:cumulant}.

\subsection{Asymmetric configurations}
\label{sec:generalBoosted}
Generalizing to an arbitrary configuration with strings that have endpoints
with transverse momentum follows naturally from combining the frameworks 
presented in Sec.~\ref{sec:generalStrings} and Sec.~\ref{sec:symmetricBoosted}.
The effects are smaller than in the symmetric configuration with opposite boosts,
since the overlap in rapidity along the $z$-axis decreases the more
transversely boosted the endpoints are. 
This results in less compression, and 
less fragmentation repulsion. We will use the configuration from
Eq.~(\ref{eq:generalConfig}), with a boost factor of $\beta = 0.1$ in opposite directions for each string.

In the right panel of ~\ref{fig:trans_average_pT}, we show the results of
boosting each string in the general configuration given by Eq.~(\ref{eq:generalConfig}) in opposite directions, then boosting back to their common
rest frame, and then performing our compression and fragmentation repulsion.
We have chosen to present the results of using $c_R = 0.4$~GeV since
larger values of this parameter are required to have visible results for
this observable.
The results are in line with our expectations from previous sections, namely
that strings with endpoints that have transverse components will compress
and repel less than strings that are completely parallel, and similarly with strings that
are not completely overlapping.

\subsection{Rotated configurations}

Configurations such as those depicted in the right-hand pane of Fig.~\ref{fig:robo} can be treated using the same arguments as for the boosted configurations. The endpoints again have non-vanishing transverse momenta, hence the rapidity spans computed along the common rapidity axis are always smaller than those in the respective string CM frames. 

In the specific example shown in ~\ref{fig:robo},  $p_{\perp,\mathrm{res}} = 0$ since each of the (1,4) and (2,3) strings have zero net $p_\perp$. Compression factors are computed from the longitudinal momentum components as in Eq.~(\ref{eq:transInternalSolution}, and the effective span taken by each hadron is projected onto the common axis using Eq.~(\ref{eq:yEffective}).

Finally, since each of the strings are at rest the $\vec{p}_{\perp,\mathrm{rel}}$ in Eq.~(\ref{eq:nT}) is zero hence the random component will dominate in the choice of azimuth direction. (A more physical choice could potentially be made by using the direction transverse to the plane spanned by the two strings, but since we consider the case of vanishing $\vec{p}_{\perp,\mathrm{rel}}$ to be of limited general interest we do not pursue this further here.)

There are many other configurations that one may consider, but with
the four configurations discussed in this work, we have presented the
overall framework for our model of fragmentation repulsion.

\section{Flow and Cumulants for Two-String Configurations}
\label{sec:cumulant}
Long-distance correlations in rapidity and azimuth have been used extensively to probe collective aspects of event structure, including flow, in both proton-proton and heavy-ion collisions. (See, e.g., \cite{Snellings:2011sz} for a succinct review of elliptic flow in heavy-ion phenomenology, and references therein.)  
Here, we focus on just one such observable,  the two-particle cumulant, $c_2\left\{2\right\}$, which is designed to suppress non-flow 
contributions. It is calculated as:
\begin{equation}
\begin{split}
    c_2\left\{2\right\} &= \Big\langle\langle e^{2i(\phi_i - \phi_j)} \rangle\Big\rangle ,\\
    &= \Bigg\langle \frac{2}{n\left(n-1\right)}\sum_{i<j}^n \cos\left(2(\phi_i - \phi_j)\right) \Bigg\rangle ,
\end{split}
\label{eq:cumulant}
\end{equation}
where in the first line the outer angle bracket is 
the average over all events, and the
inner is the average over all $n$ particles in a given event.
In the second line of Eq.~(\ref{eq:cumulant}), we have removed the
self-correlations $i = j$, and used the fact that 
the cosine function is an
even function. 

The two-particle cumulant will depend not only on the repulsion strength
$c_R$, but also on the direction of the repulsion, in particular for
cases where the strings have an overall transverse motion such as the
transversely boosted strings, where $\vec{v}_{\rm det} \neq 0$.
In this work, we will simply show the three extreme cases of the repulsion
directions for the transversely boosted configurations, as discussed in
Sec. \ref{sec:symmetricBoostedResult}

\begin{figure*}[tp]
\centering
\includegraphics[width=0.49\textwidth]{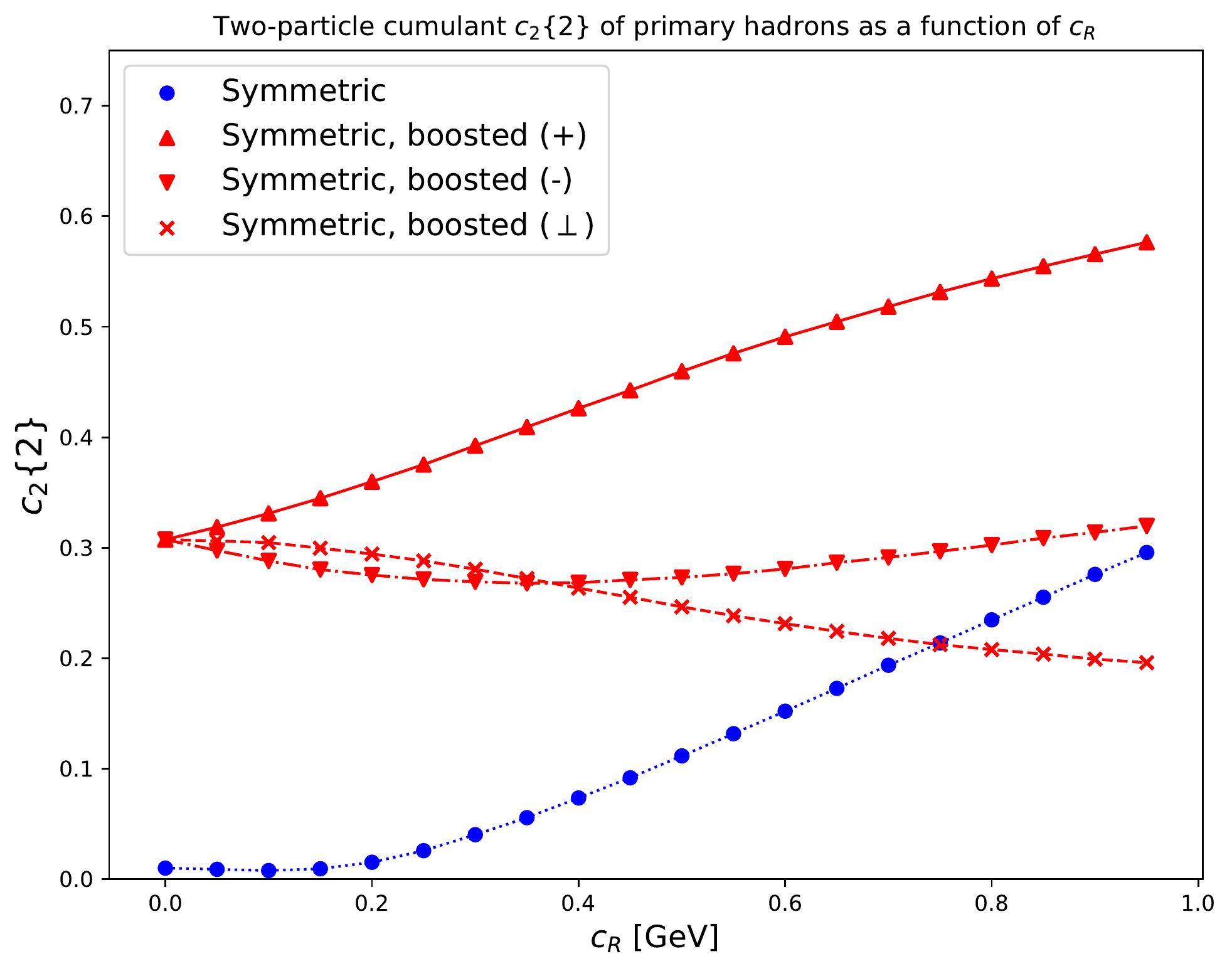}
\includegraphics[width=0.49\textwidth]{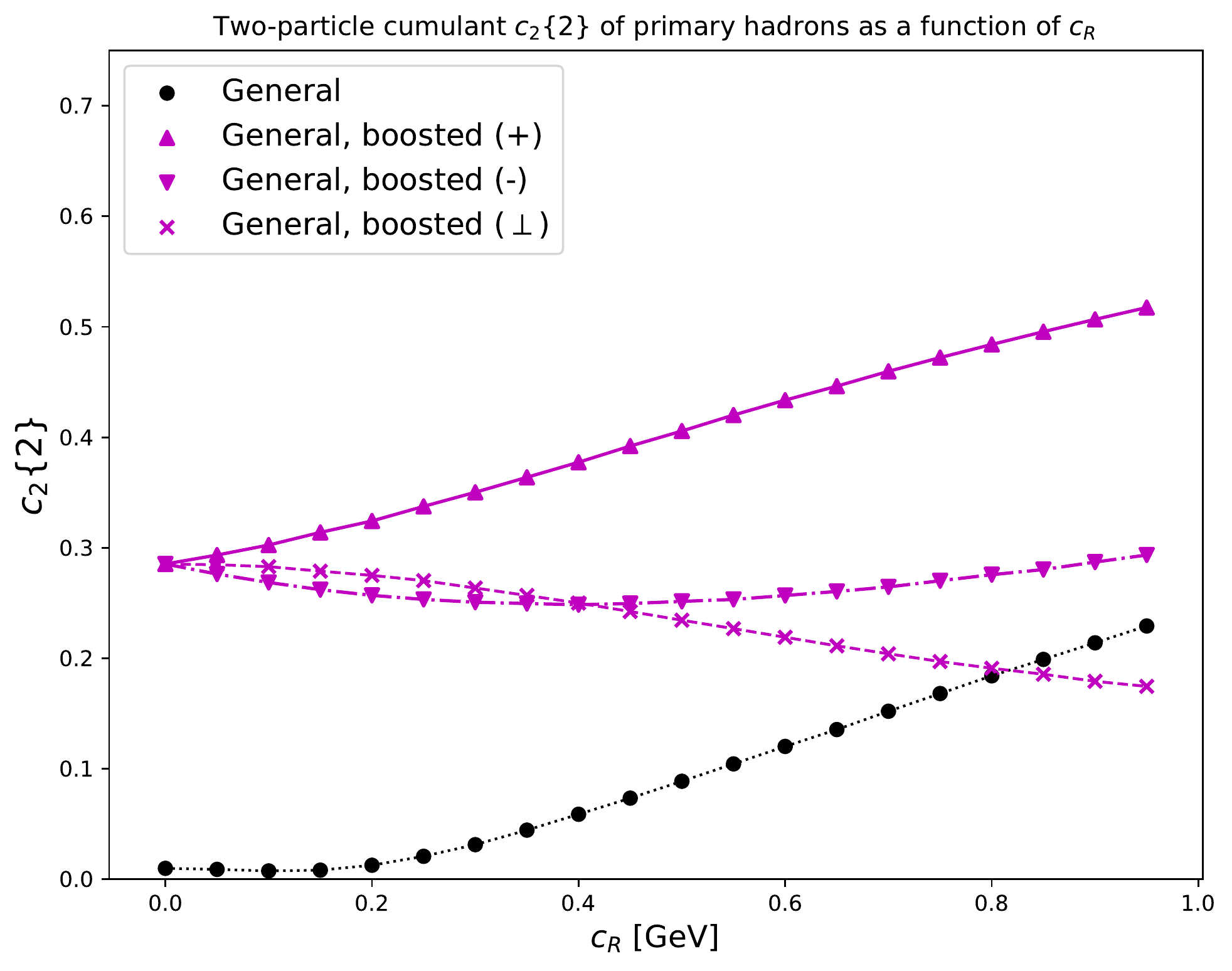}
\caption{Two-particle cumulant for the symmetric (left panel)
and the general (right panel) two-string configurations,
at the level of primary hadron production.
We show the curves for the simplest parallel two-string case, and three
variations on the equal and oppositely boosted two-string case.
The variations are: the repulsion \pT acts in the same direction as
the given string's overall transverse motion (``Boosted, $(+)$"),
the repulsion \pT acts in the opposite
direction (``Boosted, $(-)$"),
and lastly, the repulsion \pT acts perpendicularly to the 
string's boost (``Boosted, $(\perp)$''). For each curve, when $c_R = 0$, we reproduce the baseline
Lund string model.
}
\label{fig:c2}
\end{figure*}

In Fig.~\ref{fig:c2}, we plot the results for the two-particle cumulant 
for the symmetric two-string configuration 
at the level of primary hadrons, as a 
function of the repulsion constant $c_{R}$. There are four curves
in the plot. The first curve, labelled `Symmetric' is the simplest
two-string configuration, considered in Sec.~\ref{sec:simplestStrings}.
In this configuration, there is no preferred $\phi$ direction, and
it takes larger values of the repulsion constant to overcome
the Gaussian transverse momentum distribution of the Lund
fragmentation model, and to have a significant
effect on the cumulant.

The three other curves are variations on the configuration 
described in Sec.~\ref{sec:symmetricBoosted} where the two strings each
have a boost of $\beta = 0.1$ in equal and opposite directions.
The variations occur when
one adds the repulsion \pT to the primary hadrons during fragmentation.
The curves are labelled according to the direction in which the
repulsion \pT is added with respect to the given string's overall boost
direction. If we add the repulsion \pT in the same direction as the
string's motion, we can greatly enhance the two-particle cumulant.
If instead we add it in the opposite direction, we at first reduce the
two-particle cumulant, but as the repulsion gets larger, the cumulant
begins to increase.
Lastly, if we add the repulsion \pT perpendicularly to the string's motion
we greatly reduce the cumulant, but at large values of the repulsion
constant, the rate of decrease begins to level out.

We obtain analogous results for the general configuration in the right panel of
Fig.~\ref{fig:c2}, though the cumulant for all values of $c_R$
is less than for the symmetric case, due to the smaller overlap in rapidity.

We compared the symmetric parallel configuration in our fragmentation repulsion
framework to the analogous configuration in the shoving model, and found that the
two-particle cumulant is significantly smaller for the shoving model, at least
with the parameter set described in App. \ref{app:shoving}. For the shoving model,
we calculated the two-particle cumulant to be $c_2\{2\} = 0.00957$ 
(averaged over 200,000 events), which is of the order of the
baseline Lund model.

\section{Final-State Hadrons}
\label{sec:finalState}
\begin{figure}[tp]
\centering
\includegraphics[width=0.48\textwidth]{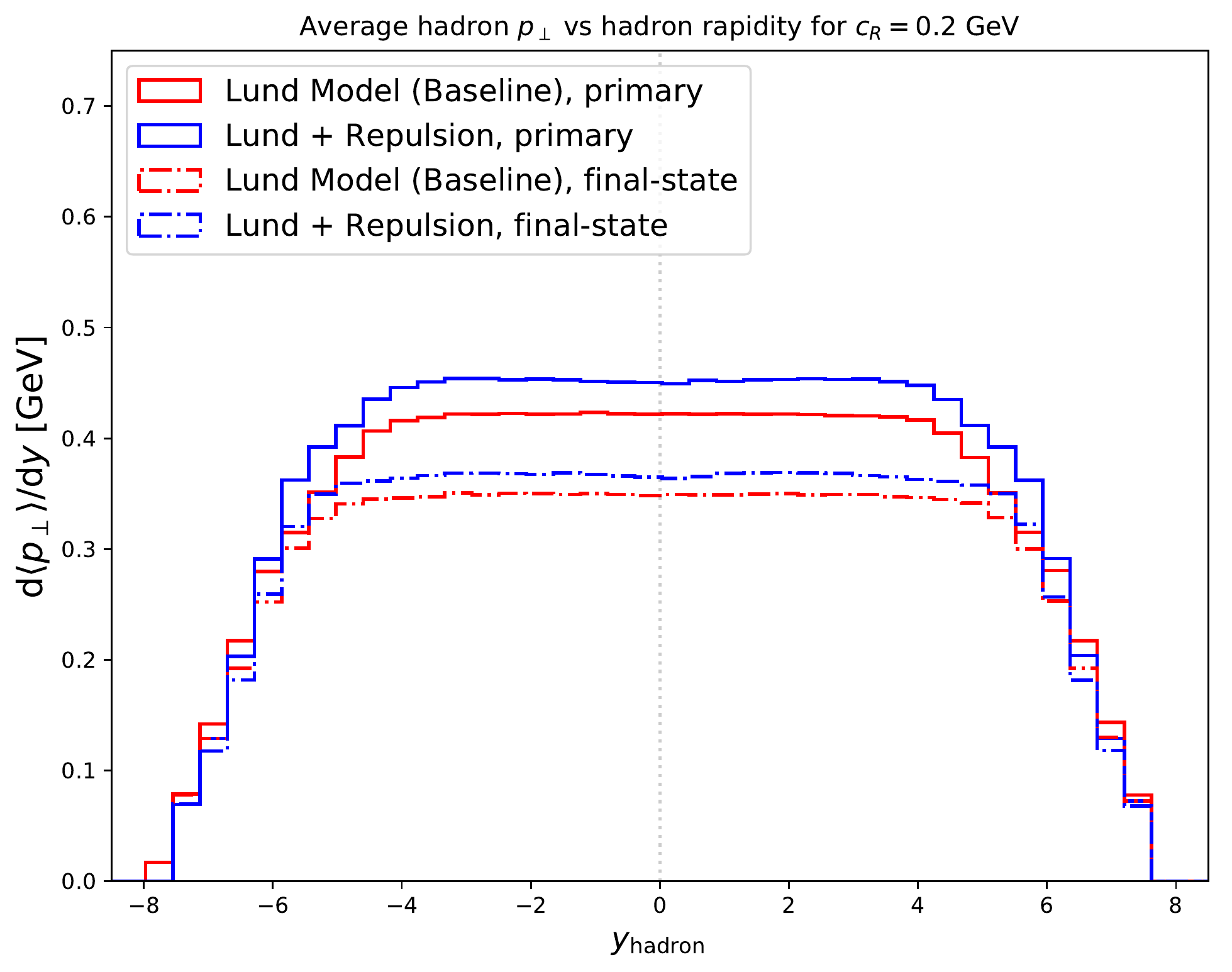}
\caption{Illustration of the net reduction of average hadron $p_\perp$ caused by allowing excited primary hadrons (solid histograms) to decay (dot-dashed histograms), for the baseline Lund model (red) and our fragmentation repulsion model (blue). The example configuration is the  symmetric parallel two-string configuration
described in Sec.~\ref{sec:simplestStrings};
the primary-hadron spectra are the same as those in Fig.~\ref{fig:plateau}.
}
\label{fig:averagepTDecays}
\end{figure}

\begin{figure*}[tp]
\includegraphics[width=0.48\textwidth]{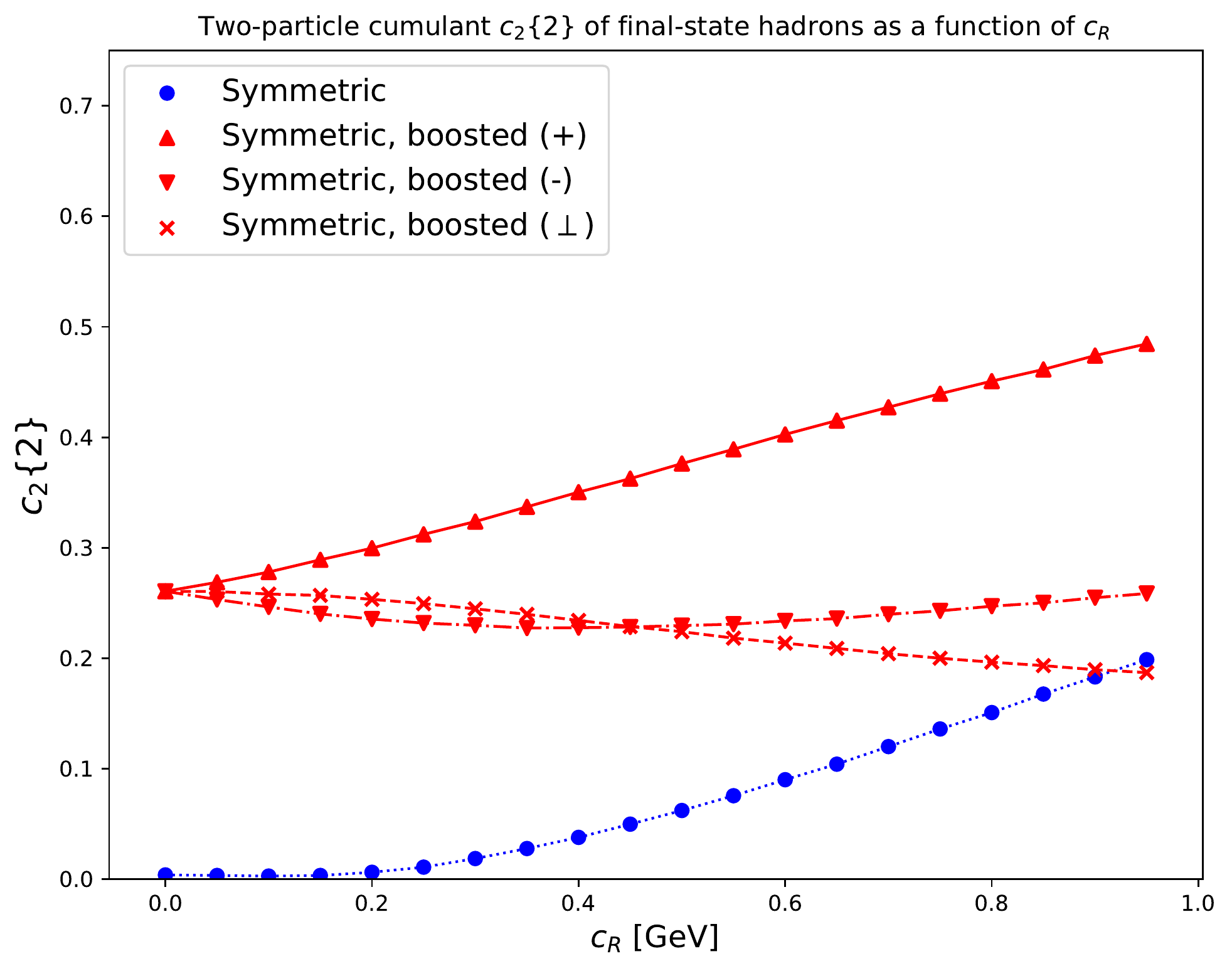}
\includegraphics[width=0.48\textwidth]{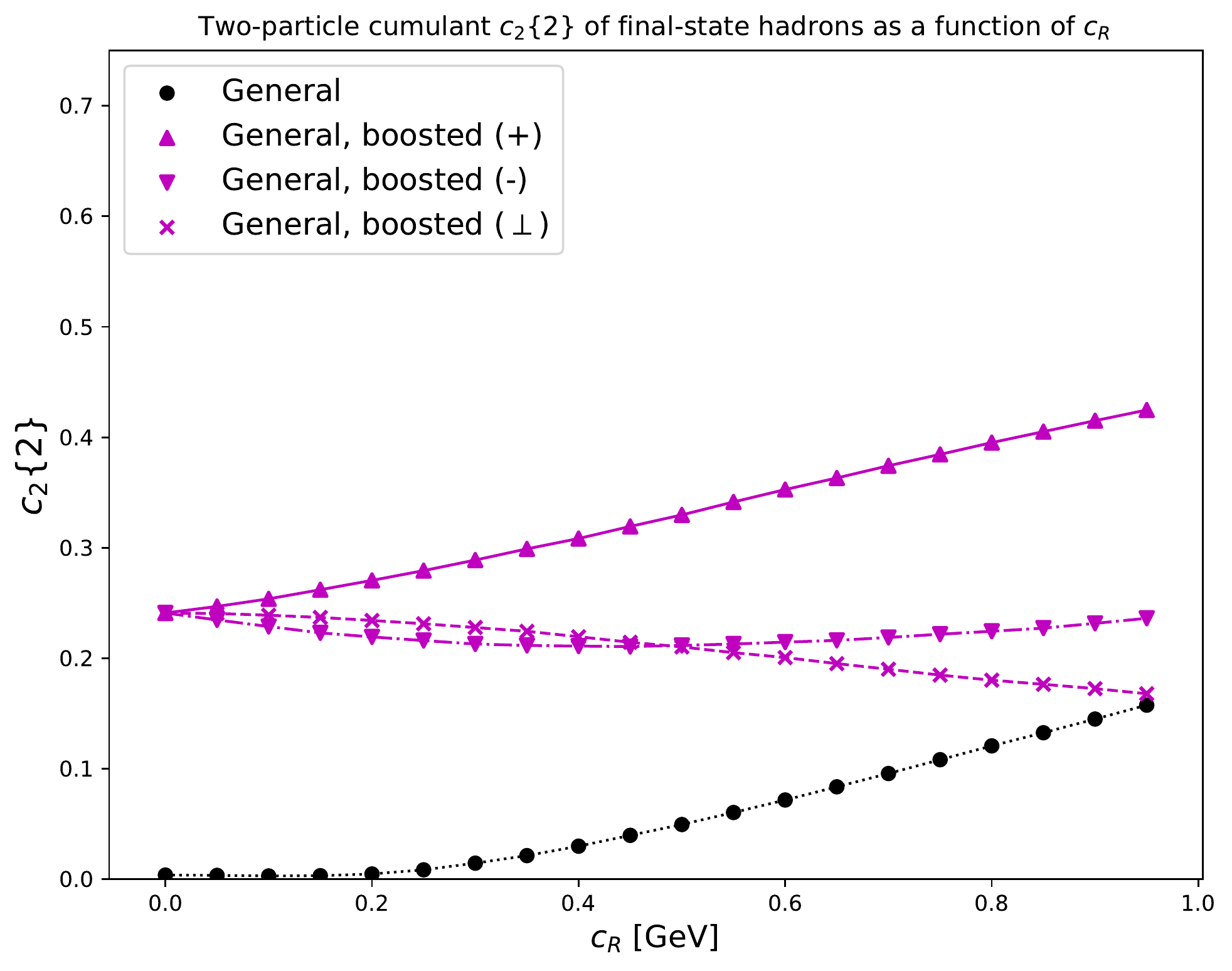}
\caption{Two-particle cumulant for final-state hadrons in the symmetric
configuration (left), and the general configuration (right), as
a function of the repulsion constant $c_R$. Both plots
exhibit the same trends as the primary hadron distributions in Fig.~\ref{fig:c2}, though the
correlations are slightly reduced, as expected from excited hadrons decaying
isotropically into potentially non-hadronic final states.
}
\label{fig:c2Decays}
\end{figure*}

In the previous sections, we considered the \pT and rapidity distributions at the level of the primary hadrons produced in the fragmentation process. Decays of those hadrons into secondaries (via processes like $\rho \to \pi\pi$, $\pi^0 \to \gamma\gamma$, etc.) will smear the distributions in rapidity and dilute the \pT enhancement per hadron.
In this section, we include  decays of all final-state particles with lifetimes shorter than $\tau = 10$~mm/$c$. 
In Pythia, this is done with the two switches:
{\tt ParticleDecays:limitTau0 = on}, and
{\tt ParticleDecays:tau0Max = 10}. With this criterion, weak\-ly decaying strange hadrons are treated as stable, while all particles with shorter lifetimes are decayed. This matches the typical definition for stable particles used at LHC.

In Fig.~\ref{fig:averagepTDecays}, we present the average hadron
\pT distribution as a function of hadron rapidity, for the 
symmetric parallel configuration from Sec.~\ref{sec:simplestStrings}.
We replot the results for the baseline Lund model (red solid) and our
fragmentation repulsion (blue solid) for primary hadrons.
Allowing excited primary hadrons to decay produces the dot-dashed lines
in Fig.~\ref{fig:averagepTDecays}, for the baseline Lund (red dot-dashed)
and our fragmentation repulsion (blue dot-dashed). 

As expected,
the plateau has been lowered for the baseline Lund model, since
excited primary hadrons can decay into non-hadronic final state
particles, which remove some of the available $p_{\perp}$.
Similarly, the fragmentation repulsion exhibits a lowering of its peak
and general structure. However, the difference between the 
structure of the fragmentation repulsion and the rapidity plateau of the
Lund model remains intact when decays are turned on, meaning our model
can still be distinguished from the baseline Lund model.

In Fig.~\ref{fig:c2Decays}, we show the effects of varying
the repulsion constant $c_R$ on the two-particle
azimuthal cumulant $c_2\{2\}$ of final-state hadrons, for the
symmetric configurations (left) and the general configurations (right).
As shown, the cumulant exhibits the same trends as the primary hadron
counterparts in Fig.~\ref{fig:c2}, though the effects have been somewhat reduced, due
to the non-hadronic particles produced during particle decays.

The key result of allowing particle decays is that our fragmentation repulsion
model, implemented at the level of the primary hadrons produced during
string fragmentation, still retains its key signatures at the level
of final-state hadrons, at least at the level of the two-string
configurations.

\section{Strings With Massive Endpoints}
\label{sec:massive}
The final generalisation we will consider in this work concerns strings with massive endpoints. 
The starting point for the compression process is the same as in the massless case, in that we rescale the 4-momenta as if the endpoints were massless:
\begin{equation}
    p^{\mu}_{\pm} \to p'^{\mu}_{\pm} = f_{\pm}p^{\mu}_{\pm} ,
    \label{eq:massiveShorten}
\end{equation}
where the subscript $\pm$ refers to the positively and negatively $z$-aligned
endpoints respectively. The compression factors are, however, slightly modified relative to those in Eq.~(\ref{eq:constraints}). 
Using the conservation of invariant mass:
\begin{equation}
    \begin{split}
        W'^2 &= W^2 - p_{\perp,R}^2 , \\
        {\rm thus}\;\; (f_+p_+ + f_-p_-)^2 &= (p_+ + p_-)^2 - p_{\perp,R}^2 ,
    \end{split}
    \label{eq:massiveConservation}
\end{equation}
where we have inserted the original and rescaled endpoint momenta in the second line.

Expanding Eq.~(\ref{eq:massiveConservation}) and rearranging gives:
\begin{equation}
    (1-f_+^2)m^2_+ + (1-f^2_-)m_-^2 + 2p_+\cdot p_- - p_{\perp,R}^2 = 2f_+f_-p_+\cdot p_- .
    \label{eq:massiveExpanded}
\end{equation}
Using the longitudinal momentum conservation to remove, e.g., $f_+$ produces a quadratic
in $f_-$ which can be simply solved to calculate the two compression factors.

After calculating the new momenta for the endpoints in the manner described above, we put the endpoints back  on shell:
\begin{equation}
    E'_{\pm} = \sqrt{m^2_{\pm} + \vec{p}'^2_{\pm}} = \sqrt{m^2_{\pm} + f^2_{\pm}\vec{p}^2_{\pm}} \leq E_{\pm} ,
    \label{eq:massiveE}
\end{equation}
where the last inequality of Eq.~(\ref{eq:massiveE}) emphasises the fact
that $f_{\pm}$ are indeed compression factors. 
With Eqs.~(\ref{eq:massiveExpanded}, \ref{eq:massiveE}), we now have a prescription for
compressing strings with massive endpoints. 
The repulsion part is the same as that 
 described in Secs. \ref{sec:fragmentationRepulsion} and \ref{sec:generalBoosted}.

\begin{figure}[tp]
\centering
\includegraphics[width=0.48\textwidth]{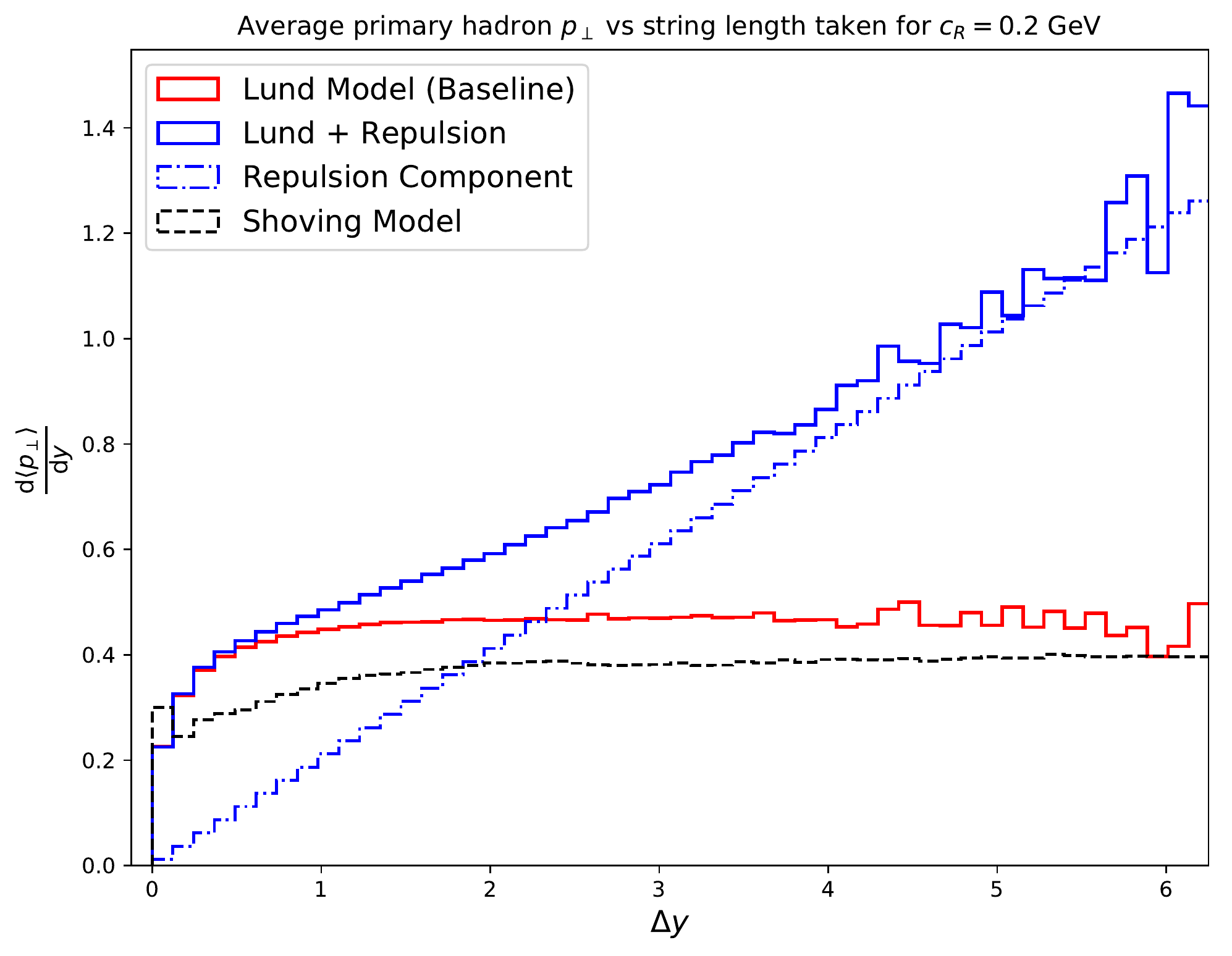}
\caption{Distribution of average hadron \pT for primary hadrons as a function of
the rapidity span of the string taken by the hadron, for the symmetric, parallel
two-string configuration with massive endpoints.
}
\label{fig:symmetricyTakenMassive}
\end{figure}

In Fig.~\ref{fig:symmetricyTakenMassive}, we present the results of our fragmentation repulsion model for the symmetric, parallel two-string configuration
with massive endpoints. 
As expected, Fig.~\ref{fig:symmetricyTakenMassive} reproduces the same characteristics
as Fig.~\ref{fig:proof}, and in particular, the significant difference between the shoving
model and the Lund model with our fragmentation repulsion remains.

Lastly, with the above prescription for handling symmetric, parallel strings 
with massive endpoints, we
can extend this formalism to the general two-string configuration using the 
frameworks of this section and Sec.~\ref{sec:generalBoosted}.
A full presentation of this and an extension to strings with gluon kinks will be
discussed in future work.

\section{Conclusion and Outlook}
\label{sec:conclusion}
We have presented a framework to compress two simple $q\bar{q}$ strings and repel them at
the level of string fragmentation, a model we call fragmentation repulsion.
We have shown that this induces an increased average \pT per hadron in regions of string overlaps and that this in turn generates non-trivial two-particle azimuthal
correlations.

With the configurations presented, one may begin to build up the more complicated
string topologies from the smaller pieces we have considered. Future work 
will look first at strings with gluon kinks, then at configurations with more than two overlapping strings. More complicated string topologies such as junctions and closed gluon loops will also need to be addressed to turn the model into a full-fledged description of 
LHC events.

A shortcoming of our work is that it does not provide a microscopic
description of the string-string interactions, unlike the shoving model. That is, we describe the effect simply in terms of an effective average \pT density that we postulate is accumulated by strings that overlap in rapidity, and which is  transferred to the hadrons that are produced in the overlapping regions. Despite its relative simplicity, the model exhibits distinctive signatures
in both average hadron transverse momentum and two-particle azimuthal
correlations which are easy to understand intuitively. 
The amount of repulsion generated via Eq.~(\ref{eq:pTrepel}) is longitudinally boost invariant, but there remains some frame dependence --- and associated ambiguities --- in our choices of rapidity and repulsion 
axes, and in the definition of the compression procedure. 
We aim to study these aspects further in future work. 

We round off by noting that, since the cluster hadronization model is based on simple $q\bar{q}$ systems not unlike those considered here, it might be possible to apply our model also in the context of the 
cluster model, to let clusters repel
off one another while losing some longitudinal momentum. However, since a cluster
undergoes fissioning and decay, the repulsion would need to be split between the two products in the respective processes.

\section*{Acknowledgements}
We thank G\"{o}sta Gustafson for his valuable comments on the work, Christian Bierlich for his helpful comments about the current implementation of the shoving model in Pythia 8.2, and Helen Brooks \& Johannes Bellm for discussions of our work. CBD is supported by the Australian Government Research Training 
Program Scholarship and the J.~L.~William Scholarship. This work was funded in part by the Australian Research Council via Discovery Project DP170100708 -- “Emergent Phenomena in Quantum Chromodynamics” -- and in part by the European Union’s Horizon 2020 research and innovation programme under the Marie Sk\l{}odowska-Curie grant agreement No 722104 -- MCnetITN3.  

\bibliography{main}

\begin{thebibliography}{10}
\providecommand{\url}[1]{\texttt{#1}}
\providecommand{\urlprefix}{URL }
\expandafter\ifx\csname urlstyle\endcsname\relax
  \providecommand{\doi}[1]{doi:\discretionary{}{}{}#1}\else
  \providecommand{\doi}{doi:\discretionary{}{}{}\begingroup
  \urlstyle{rm}\Url}\fi
\providecommand{\eprint}[2][]{\url{#2}}

\bibitem{Andersson:1983ia}
B.~Andersson, G.~Gustafson, G.~Ingelman and T.~Sj{\"o}strand,
\newblock \emph{{Parton Fragmentation and String Dynamics}},
\newblock Phys. Rept. \textbf{97}, 31 (1983),
\newblock \doi{10.1016/0370-1573(83)90080-7}.

\bibitem{Andersson:1983jt}
B.~Andersson, G.~Gustafson and B.~Soderberg,
\newblock \emph{{A General Model for Jet Fragmentation}},
\newblock Z. Phys. \textbf{C20}, 317 (1983),
\newblock \doi{10.1007/BF01407824}.

\bibitem{Andersson:1998tv}
B.~Andersson,
\newblock \emph{{The Lund model}},
\newblock Camb. Monogr. Part. Phys. Nucl. Phys. Cosmol. \textbf{7}, 1 (1997).

\bibitem{SJOSTRAND1984469}
T.~Sjöstrand,
\newblock \emph{Jet fragmentation of multiparton configurations in a string
  framework},
\newblock Nuclear Physics B \textbf{248}(2), 469  (1984),
\newblock \doi{https://doi.org/10.1016/0550-3213(84)90607-2}.

\bibitem{FIELD198365}
R.~D. Field and S.~Wolfram,
\newblock \emph{{A QCD model for $e^+e^-$ annihilation}},
\newblock Nuclear Physics B \textbf{213}(1), 65  (1983),
\newblock \doi{https://doi.org/10.1016/0550-3213(83)90175-X}.

\bibitem{GOTTSCHALK1984349}
T.~D. Gottschalk,
\newblock \emph{{An improved description of hadronization in the QCD cluster
  model for $e^+e^{-}$ annihilation}},
\newblock Nuclear Physics B \textbf{239}(2), 349  (1984),
\newblock \doi{https://doi.org/10.1016/0550-3213(84)90253-0}.

\bibitem{WEBBER1984492}
B.~Webber,
\newblock \emph{{A QCD model for jet fragmentation including soft gluon
  interference}},
\newblock Nuclear Physics B \textbf{238}(3), 492  (1984),
\newblock \doi{http://dx.doi.org/10.1016/0550-3213(84)90333-X}.

\bibitem{Sjostrand:2006za}
T.~Sj{\"o}strand, S.~Mrenna and P.~Z. Skands,
\newblock \emph{{PYTHIA 6.4 Physics and Manual}},
\newblock JHEP \textbf{05}, 026 (2006),
\newblock \doi{10.1088/1126-6708/2006/05/026},
\newblock \eprint{hep-ph/0603175}.

\bibitem{Sjostrand:2007gs}
T.~Sj{\"o}strand, S.~Mrenna and P.~Z. Skands,
\newblock \emph{{A Brief Introduction to PYTHIA 8.1}},
\newblock Comput. Phys. Commun. \textbf{178}, 852 (2008),
\newblock \doi{10.1016/j.cpc.2008.01.036},
\newblock \eprint{0710.3820}.

\bibitem{Sjostrand:2014zea}
T.~Sj{\"o}strand, S.~Ask, J.~R. Christiansen, R.~Corke, N.~Desai, P.~Ilten,
  S.~Mrenna, S.~Prestel, C.~O. Rasmussen and P.~Z. Skands,
\newblock \emph{{An Introduction to PYTHIA 8.2}},
\newblock Comput. Phys. Commun. \textbf{191}, 159 (2015),
\newblock \doi{10.1016/j.cpc.2015.01.024},
\newblock \eprint{1410.3012}.

\bibitem{Werner:2005jf}
K.~Werner, F.-M. Liu and T.~Pierog,
\newblock \emph{{Parton ladder splitting and the rapidity dependence of
  transverse momentum spectra in deuteron-gold collisions at RHIC}},
\newblock Phys. Rev. \textbf{C74}, 044902 (2006),
\newblock \doi{10.1103/PhysRevC.74.044902},
\newblock \eprint{hep-ph/0506232}.

\bibitem{Pierog:2013ria}
T.~Pierog, I.~Karpenko, J.~M. Katzy, E.~Yatsenko and K.~Werner,
\newblock \emph{{EPOS LHC: Test of collective hadronization with data measured
  at the CERN Large Hadron Collider}},
\newblock Phys. Rev. \textbf{C92}(3), 034906 (2015),
\newblock \doi{10.1103/PhysRevC.92.034906},
\newblock \eprint{1306.0121}.

\bibitem{Bahr:2008pv}
M.~B{\"a}hr \emph{et~al.},
\newblock \emph{{Herwig++ Physics and Manual}},
\newblock Eur. Phys. J. \textbf{C58}, 639 (2008),
\newblock \doi{10.1140/epjc/s10052-008-0798-9},
\newblock \eprint{0803.0883}.

\bibitem{Bellm:2015jjp}
J.~Bellm \emph{et~al.},
\newblock \emph{{Herwig 7.0/Herwig++ 3.0 release note}},
\newblock Eur. Phys. J. \textbf{C76}(4), 196 (2016),
\newblock \doi{10.1140/epjc/s10052-016-4018-8},
\newblock \eprint{1512.01178}.

\bibitem{Bellm:2017bvx}
J.~Bellm \emph{et~al.},
\newblock \emph{{Herwig 7.1 Release Note}}  (2017),
\newblock \eprint{1705.06919}.

\bibitem{Gleisberg:2008ta}
T.~Gleisberg, S.~Hoeche, F.~Krauss, M.~Sch{\"o}nherr, S.~Schumann, F.~Siegert
  and J.~Winter,
\newblock \emph{{Event generation with SHERPA 1.1}},
\newblock JHEP \textbf{02}, 007 (2009),
\newblock \doi{10.1088/1126-6708/2009/02/007},
\newblock \eprint{0811.4622}.

\bibitem{Bothmann:2019yzt}
E.~Bothmann \emph{et~al.},
\newblock \emph{{Event Generation with Sherpa 2.2}},
\newblock SciPost Phys. \textbf{7}, 034 (2019),
\newblock \doi{10.21468/SciPostPhys.7.3.034},
\newblock \eprint{1905.09127}.

\bibitem{Karneyeu:2013aha}
A.~Karneyeu, L.~Mijovic, S.~Prestel and P.~Z. Skands,
\newblock \emph{{MCPLOTS: a particle physics resource based on volunteer
  computing}},
\newblock Eur. Phys. J. \textbf{C74}, 2714 (2014),
\newblock \doi{10.1140/epjc/s10052-014-2714-9},
\newblock \eprint{1306.3436}.

\bibitem{Adam:2015qaa}
J.~Adam \emph{et~al.},
\newblock \emph{{Measurement of pion, kaon and proton production in
  proton–proton collisions at $\sqrt{s} = 7$ TeV}},
\newblock Eur. Phys. J. \textbf{C75}(5), 226 (2015),
\newblock \doi{10.1140/epjc/s10052-015-3422-9},
\newblock \eprint{1504.00024}.

\bibitem{Acharya:2018orn}
S.~Acharya \emph{et~al.},
\newblock \emph{{Multiplicity dependence of light-flavor hadron production in
  pp collisions at $\sqrt{s}$ = 7 TeV}},
\newblock Phys. Rev. \textbf{C99}(2), 024906 (2019),
\newblock \doi{10.1103/PhysRevC.99.024906},
\newblock \eprint{1807.11321}.

\bibitem{Khachatryan:2010gv}
V.~Khachatryan \emph{et~al.},
\newblock \emph{{Observation of Long-Range Near-Side Angular Correlations in
  Proton-Proton Collisions at the LHC}},
\newblock JHEP \textbf{09}, 091 (2010),
\newblock \doi{10.1007/JHEP09(2010)091},
\newblock \eprint{1009.4122}.

\bibitem{Velicanu:2011zz}
D.~Velicanu,
\newblock \emph{{Ridge correlation structure in high multiplicity pp collisions
  with CMS}},
\newblock J. Phys. \textbf{G38}, 124051 (2011),
\newblock \doi{10.1088/0954-3899/38/12/124051},
\newblock \eprint{1107.2196}.

\bibitem{Aad:2015gqa}
G.~Aad \emph{et~al.},
\newblock \emph{{Observation of Long-Range Elliptic Azimuthal Anisotropies in
  $\sqrt{s}=$13 and 2.76 TeV $pp$ Collisions with the ATLAS Detector}},
\newblock Phys. Rev. Lett. \textbf{116}(17), 172301 (2016),
\newblock \doi{10.1103/PhysRevLett.116.172301},
\newblock \eprint{1509.04776}.

\bibitem{Khachatryan:2015lva}
V.~Khachatryan \emph{et~al.},
\newblock \emph{{Measurement of long-range near-side two-particle angular
  correlations in pp collisions at $\sqrt s =$13 TeV}},
\newblock Phys. Rev. Lett. \textbf{116}(17), 172302 (2016),
\newblock \doi{10.1103/PhysRevLett.116.172302},
\newblock \eprint{1510.03068}.

\bibitem{Khachatryan:2016txc}
V.~Khachatryan \emph{et~al.},
\newblock \emph{{Evidence for collectivity in pp collisions at the LHC}},
\newblock Phys. Lett. \textbf{B765}, 193 (2017),
\newblock \doi{10.1016/j.physletb.2016.12.009},
\newblock \eprint{1606.06198}.

\bibitem{Aaboud:2016yar}
M.~Aaboud \emph{et~al.},
\newblock \emph{{Measurements of long-range azimuthal anisotropies and
  associated Fourier coefficients for $pp$ collisions at $\sqrt{s}=5.02$ and
  $13$ TeV and $p$+Pb collisions at $\sqrt{s_{\mathrm{NN}}}=5.02$ TeV with the
  ATLAS detector}},
\newblock Phys. Rev. \textbf{C96}(2), 024908 (2017),
\newblock \doi{10.1103/PhysRevC.96.024908},
\newblock \eprint{1609.06213}.

\bibitem{Aaboud:2017acw}
M.~Aaboud \emph{et~al.},
\newblock \emph{{Measurement of multi-particle azimuthal correlations in $pp$,
  $p+$Pb and low-multiplicity Pb$+$Pb collisions with the ATLAS detector}},
\newblock Eur. Phys. J. \textbf{C77}(6), 428 (2017),
\newblock \doi{10.1140/epjc/s10052-017-4988-1},
\newblock \eprint{1705.04176}.

\bibitem{Chatrchyan:2013qsa}
S.~Chatrchyan \emph{et~al.},
\newblock \emph{{Measurement of Neutral Strange Particle Production in the
  Underlying Event in Proton-Proton Collisions at $\sqrt{s}$ = 7 TeV}},
\newblock Phys. Rev. \textbf{D88}, 052001 (2013),
\newblock \doi{10.1103/PhysRevD.88.052001},
\newblock \eprint{1305.6016}.

\bibitem{Aad:2019xek}
G.~Aad \emph{et~al.},
\newblock \emph{{Measurement of $K_S^0$ and $\Lambda^0$ production in $t
  \bar{t}$ dileptonic events in $pp$ collisions at $\sqrt{s} =$ 7 TeV with the
  ATLAS detector}}  (2019),
\newblock \eprint{1907.10862}.

\bibitem{Cui:2019jbt}
P.~Cui,
\newblock \emph{{Strangeness Production in Jets and Underlying Events in pp
  Collisions at $\sqrt{s}$ = 13 TeV with ALICE}}  (2019),
\newblock \eprint{1909.05471}.

\bibitem{Sjostrand:2013cya}
T.~Sjöstrand,
\newblock \emph{{Colour reconnection and its effects on precise measurements at
  the LHC}} (2013), \eprint{1310.8073}.

\bibitem{Sjostrand:2017ele}
T.~Sj{\"o}strand,
\newblock \emph{{Colour Reconnections from LEP to Future Colliders}},
\newblock In \emph{{Proceedings, Parton Radiation and Fragmentation from LHC to
  FCC-ee: CERN, Geneva, Switzerland, November 22-23, 2016, arXiv:1702.01329}},
  pp. 144--148 (2017).

\bibitem{Bierlich:2014xba}
C.~Bierlich, G.~Gustafson, L.~L{\"o}nnblad and A.~Tarasov,
\newblock \emph{{Effects of Overlapping Strings in pp Collisions}},
\newblock JHEP \textbf{03}, 148 (2015),
\newblock \doi{10.1007/JHEP03(2015)148},
\newblock \eprint{1412.6259}.

\bibitem{Bierlich:2015rha}
C.~Bierlich and J.~R. Christiansen,
\newblock \emph{{Effects of color reconnection on hadron flavor observables}},
\newblock Phys. Rev. \textbf{D92}(9), 094010 (2015),
\newblock \doi{10.1103/PhysRevD.92.094010},
\newblock \eprint{1507.02091}.

\bibitem{AMBJORN1984533}
J.~Ambjørn, P.~Olesen and C.~Peterson,
\newblock \emph{Stochastic confinement and dimensional reduction (ii).
  three-dimensional su(2) lattice gauge theory},
\newblock Nuclear Physics B \textbf{240}(4), 533  (1984),
\newblock \doi{https://doi.org/10.1016/0550-3213(84)90242-6}.

\bibitem{Bali:2000un}
G.~S. Bali,
\newblock \emph{{Casimir scaling of SU(3) static potentials}},
\newblock Phys. Rev. \textbf{D62}, 114503 (2000),
\newblock \doi{10.1103/PhysRevD.62.114503},
\newblock \eprint{hep-lat/0006022}.

\bibitem{Bierlich:2016vgw}
C.~Bierlich, G.~Gustafson and L.~L{\"o}nnblad,
\newblock \emph{{A shoving model for collectivity in hadronic collisions}}
  (2016),
\newblock \eprint{1612.05132}.

\bibitem{Bierlich:2017vhg}
C.~Bierlich, G.~Gustafson and L.~Lönnblad,
\newblock \emph{{Collectivity without plasma in hadronic collisions}},
\newblock Phys. Lett. \textbf{B779}, 58 (2018),
\newblock \doi{10.1016/j.physletb.2018.01.069},
\newblock \eprint{1710.09725}.

\bibitem{Flensburg:2011kk}
C.~Flensburg, G.~Gustafson and L.~L{\"o}nnblad,
\newblock \emph{{Inclusive and Exclusive Observables from Dipoles in High
  Energy Collisions}},
\newblock JHEP \textbf{08}, 103 (2011),
\newblock \doi{10.1007/JHEP08(2011)103},
\newblock \eprint{1103.4321}.

\bibitem{Bierlich:2018xfw}
C.~Bierlich, G.~Gustafson, L.~Lönnblad and H.~Shah,
\newblock \emph{{The Angantyr model for Heavy-Ion Collisions in PYTHIA8}},
\newblock JHEP \textbf{10}, 134 (2018),
\newblock \doi{10.1007/JHEP10(2018)134},
\newblock \eprint{1806.10820}.

\bibitem{Ortiz:2013yxa}
A.~Ortiz~Velasquez, P.~Christiansen, E.~Cuautle~Flores, I.~Maldonado~Cervantes
  and G.~Paić,
\newblock \emph{{Color Reconnection and Flowlike Patterns in $pp$ Collisions}},
\newblock Phys. Rev. Lett. \textbf{111}(4), 042001 (2013),
\newblock \doi{10.1103/PhysRevLett.111.042001},
\newblock \eprint{1303.6326}.

\bibitem{Christiansen:2015yqa}
J.~R. Christiansen and P.~Z. Skands,
\newblock \emph{{String Formation Beyond Leading Colour}},
\newblock JHEP \textbf{08}, 003 (2015),
\newblock \doi{10.1007/JHEP08(2015)003},
\newblock \eprint{1505.01681}.

\bibitem{Fischer:2016zzs}
N.~Fischer and T.~Sjöstrand,
\newblock \emph{{Thermodynamical String Fragmentation}},
\newblock JHEP \textbf{01}, 140 (2017),
\newblock \doi{10.1007/JHEP01(2017)140},
\newblock \eprint{1610.09818}.

\bibitem{Gieseke:2018gff}
S.~Gieseke, P.~Kirchgaeßer, S.~Pl{\"a}tzer and A.~Siodmok,
\newblock \emph{{Colour Reconnection from Soft Gluon Evolution}},
\newblock JHEP \textbf{11}, 149 (2018),
\newblock \doi{10.1007/JHEP11(2018)149},
\newblock \eprint{1808.06770}.

\bibitem{Duncan:2018gfk}
C.~B. Duncan and P.~Kirchgaeßer,
\newblock \emph{{Kinematic strangeness production in cluster hadronization}},
\newblock Eur. Phys. J. \textbf{C79}(1), 61 (2019),
\newblock \doi{10.1140/epjc/s10052-019-6573-2},
\newblock \eprint{1811.10336}.

\bibitem{Celik:1980td}
T.~Celik, F.~Karsch and H.~Satz,
\newblock \emph{{A PERCOLATION APPROACH TO STRONGLY INTERACTING MATTER}},
\newblock Phys. Lett. B \textbf{97}, 128 (1980),
\newblock \doi{10.1016/0370-2693(80)90564-X}.

\bibitem{Ferreiro:2003dw}
E.~Ferreiro, F.~del Moral and C.~Pajares,
\newblock \emph{{Transverse momentum fluctuations and percolation of strings}},
\newblock Phys. Rev. C \textbf{69}, 034901 (2004),
\newblock \doi{10.1103/PhysRevC.69.034901},
\newblock \eprint{hep-ph/0303137}.

\bibitem{Braun:2015eoa}
M.~Braun, J.~Dias~de Deus, A.~Hirsch, C.~Pajares, R.~Scharenberg and
  B.~Srivastava,
\newblock \emph{{De-Confinement and Clustering of Color Sources in Nuclear
  Collisions}},
\newblock Phys. Rept. \textbf{599}, 1 (2015),
\newblock \doi{10.1016/j.physrep.2015.09.003},
\newblock \eprint{1501.01524}.

\bibitem{Bautista:2015kwa}
I.~Bautista, A.~F. Téllez and P.~Ghosh,
\newblock \emph{{Indication of change of phase in high-multiplicity
  proton-proton events at LHC in String Percolation Model}},
\newblock Phys. Rev. D \textbf{92}(7), 071504 (2015),
\newblock \doi{10.1103/PhysRevD.92.071504},
\newblock \eprint{1509.02278}.

\bibitem{Ramirez:2017oef}
J.~Ramírez, A.~Fernández~Téllez and I.~Bautista,
\newblock \emph{{String percolation threshold for elliptically bounded
  systems}},
\newblock Physica A \textbf{488}, 8 (2017),
\newblock \doi{10.1016/j.physa.2017.07.002},
\newblock \eprint{1707.06395}.

\bibitem{Blok:2018xes}
B.~Blok and U.~A. Wiedemann,
\newblock \emph{{Collectivity in pp from resummed interference effects?}},
\newblock Phys. Lett. B \textbf{795}, 259 (2019),
\newblock \doi{10.1016/j.physletb.2019.05.038},
\newblock \eprint{1812.04113}.

\bibitem{Levin:2011fb}
E.~Levin and A.~H. Rezaeian,
\newblock \emph{{The Ridge from the BFKL evolution and beyond}},
\newblock Phys. Rev. D \textbf{84}, 034031 (2011),
\newblock \doi{10.1103/PhysRevD.84.034031},
\newblock \eprint{1105.3275}.

\bibitem{Bzdak:2013zma}
A.~Bzdak, B.~Schenke, P.~Tribedy and R.~Venugopalan,
\newblock \emph{{Initial state geometry and the role of hydrodynamics in
  proton-proton, proton-nucleus and deuteron-nucleus collisions}},
\newblock Phys. Rev. C \textbf{87}(6), 064906 (2013),
\newblock \doi{10.1103/PhysRevC.87.064906},
\newblock \eprint{1304.3403}.

\bibitem{Yan:2014nsa}
L.~Yan, J.-Y. Ollitrault and A.~M. Poskanzer,
\newblock \emph{{Azimuthal Anisotropy Distributions in High-Energy
  Collisions}},
\newblock Phys. Lett. B \textbf{742}, 290 (2015),
\newblock \doi{10.1016/j.physletb.2015.01.039},
\newblock \eprint{1408.0921}.

\bibitem{Albacete:2016gxu}
J.~L. Albacete, H.~Petersen and A.~Soto-Ontoso,
\newblock \emph{{Correlated wounded hot spots in proton-proton interactions}},
\newblock Phys. Rev. C \textbf{95}(6), 064909 (2017),
\newblock \doi{10.1103/PhysRevC.95.064909},
\newblock \eprint{1612.06274}.

\bibitem{Dumitru:2010iy}
A.~Dumitru, K.~Dusling, F.~Gelis, J.~Jalilian-Marian, T.~Lappi and
  R.~Venugopalan,
\newblock \emph{{The Ridge in proton-proton collisions at the LHC}},
\newblock Phys. Lett. B \textbf{697}, 21 (2011),
\newblock \doi{10.1016/j.physletb.2011.01.024},
\newblock \eprint{1009.5295}.

\bibitem{Dusling:2012iga}
K.~Dusling and R.~Venugopalan,
\newblock \emph{{Azimuthal collimation of long range rapidity correlations by
  strong color fields in high multiplicity hadron-hadron collisions}},
\newblock Phys. Rev. Lett. \textbf{108}, 262001 (2012),
\newblock \doi{10.1103/PhysRevLett.108.262001},
\newblock \eprint{1201.2658}.

\bibitem{Kurkela:2018qeb}
A.~Kurkela, U.~A. Wiedemann and B.~Wu,
\newblock \emph{{Opacity dependence of elliptic flow in kinetic theory}},
\newblock Eur. Phys. J. C \textbf{79}(9), 759 (2019),
\newblock \doi{10.1140/epjc/s10052-019-7262-x},
\newblock \eprint{1805.04081}.

\bibitem{Schlichting:2016sqo}
S.~Schlichting and P.~Tribedy,
\newblock \emph{{Collectivity in Small Collision Systems: An Initial-State
  Perspective}},
\newblock Adv. High Energy Phys. \textbf{2016}, 8460349 (2016),
\newblock \doi{10.1155/2016/8460349},
\newblock \eprint{1611.00329}.

\bibitem{Bierlich:2020kzy}
C.~Bierlich,
\newblock \emph{{Sources of multiparticle correlations: a microscopic
  perspective}},
\newblock In \emph{{49th International Symposium on Multiparticle Dynamics}}
  (2020), \eprint{2002.10746}.

\bibitem{Ferreres-Sole:2018vgo}
S.~Ferreres-Solé and T.~Sj{\"o}strand,
\newblock \emph{{The space–time structure of hadronization in the Lund
  model}},
\newblock Eur. Phys. J. \textbf{C78}(11), 983 (2018),
\newblock \doi{10.1140/epjc/s10052-018-6459-8},
\newblock \eprint{1808.04619}.

\bibitem{Schwinger:1951nm}
J.~S. Schwinger,
\newblock \emph{{On gauge invariance and vacuum polarization}},
\newblock Phys. Rev. \textbf{82}, 664 (1951),
\newblock \doi{10.1103/PhysRev.82.664}.

\bibitem{Skands:2014pea}
P.~Skands, S.~Carrazza and J.~Rojo,
\newblock \emph{{Tuning PYTHIA 8.1: the Monash 2013 Tune}},
\newblock Eur. Phys. J. \textbf{C74}(8), 3024 (2014),
\newblock \doi{10.1140/epjc/s10052-014-3024-y},
\newblock \eprint{1404.5630}.

\bibitem{Amoroso:2018qga}
S.~Amoroso, S.~Caron, A.~Jueid, R.~Ruiz~de Austri and P.~Skands,
\newblock \emph{{Estimating QCD uncertainties in Monte Carlo event generators
  for gamma-ray dark matter searches}},
\newblock JCAP \textbf{1905}(05), 007 (2019),
\newblock \doi{10.1088/1475-7516/2019/05/007},
\newblock \eprint{1812.07424}.

\bibitem{PhysRevB.3.3821}
L.~Kramer,
\newblock \emph{{Thermodynamic Behavior of Type-II Superconductors with Small
  $\ensuremath{\kappa}$ near the Lower Critical Field}},
\newblock Phys. Rev. B \textbf{3}, 3821 (1971),
\newblock \doi{10.1103/PhysRevB.3.3821}.

\bibitem{PhysRevB.83.054516}
A.~Chaves, F.~M. Peeters, G.~A. Farias and M.~V. Milo\ifmmode \check{s}\else
  \v{s}\fi{}evi\ifmmode~\acute{c}\else \'{c}\fi{},
\newblock \emph{Vortex-vortex interaction in bulk superconductors:
  Ginzburg-landau theory},
\newblock Phys. Rev. B \textbf{83}, 054516 (2011),
\newblock \doi{10.1103/PhysRevB.83.054516}.

\bibitem{Bellm:2019wrh}
J.~Bellm, C.~B. Duncan, S.~Gieseke, M.~Myska and A.~Si{\'o}dmok,
\newblock \emph{{Spacetime colour reconnection in Herwig 7}},
\newblock Eur. Phys. J. \textbf{C79}(12), 1003 (2019),
\newblock \doi{10.1140/epjc/s10052-019-7533-6},
\newblock \eprint{1909.08850}.

\bibitem{Snellings:2011sz}
R.~Snellings,
\newblock \emph{{Elliptic Flow: A Brief Review}},
\newblock New J. Phys. \textbf{13}, 055008 (2011),
\newblock \doi{10.1088/1367-2630/13/5/055008},
\newblock \eprint{1102.3010}.

\end{thebibliography}

\appendix

\section{String Fragmentation in Pythia}
\label{app:fragmentationDetails}
The Lund string fragments probabilistically by taking steps along
the lightcone momenta $W_{\pm}$, where $W_+W_- = W^2$.
A hadron is created by taking a fraction $z_h$ of a given end's lightcone momentum
and the rest of the string keeps $1-z_h$ of the lightcone momentum. In order to put the
hadron on shell, we also need to take some lightcone momentum from the other end.

Following the notational convention of \cite{SJOSTRAND1984469},
if we have taken $i$ iterative steps in the fragmentation process, producing
$q_i\qb_i$ pairs, each of which take a fraction $z_i$, and $0\leq z_i \leq 1$, we can
write the fractions of the initial total $W_{\pm}$ taken at each step:
\begin{equation}
\begin{split}
    x_{+,i} &= z_i \prod_{j=1}^{i-1}\left(1-z_j\right), \\
    \mathrm{and} \;\; x_{-,i} &= \frac{m_{\perp,i}^2}{x_{+,i}W^2} ,\\
    \mathrm{since} \;\; m_{\perp,i}^2 &= x_{+,i}x_{-,i}W^2 ,
\end{split}
\end{equation}
where we have assumed without loss of generality that the hadrons have been
fragmenting from the $W_+$ end of the string.

Since the string can fragment from either end of the string, Pythia 
needs two sets of
these $x_{\pm}$ pairs, where now the $\pm$ sign refers to the lightcone momenta
of the opposite end of the given fragmenting end. We will label them $x$ and $\tilde{x}$.
These two pairs track how much has been taken from the two end points in the
two different directions, and the differences are the amount of lightcone momentum
actually left:
\begin{equation}
    \bar{x}_{\mathrm{tot}, +} = x_{+} - \tilde{x}_- , \mathrm{and } \;\bar{x}_{\mathrm{tot}, -} = \tilde{x}_+ - x_-, 
\end{equation}

Using Eq.~(\ref{eq:simpleRap}), we can now calculate the rapidity span
of the string that a fragmenting hadron $i$ takes with it:
\begin{equation}
    \Delta y = \ln\left(\frac{\bar{x}_{\mathrm{tot},+}\bar{x}_{\mathrm{tot},-}}{(\bar{x}_{\mathrm{tot},+}-x_{h,+})(\bar{x}_{\mathrm{tot},-} - x_{h,-})}\right) ,
    \label{eq:hadronspan}
\end{equation}
where $x_{h,\pm}$ is the lightcone momentum fraction taken by a new hadron fragmentation
from the positive end and negative end respectively.

At some cutoff invariant mass $W^2_{\mathrm{stop}}$, this fragmentation
process stops, and the remnant string is broken into two
final hadrons.

\section{Shoving Model Parameters}
\label{app:shoving}
In the shoving model (as implemented in Pythia 8.2), there are several 
parameters that govern the rate
and amount of shoving. We summarise the parameter values we used
to produce Fig.~\ref{fig:plateau} in Tab.~\ref{table:shoving}. We did not
include the flavour changing aspects of the Rope model.

\begin{table}[ht]
\centering
\begin{tabular}{||l c ||} 
 \hline
 Parameter & Value  \\ [0.5ex] 
 \hline\hline
 {\tt Ropewalk:rCutOff} & 10.0  \\ 
 {\tt Ropewalk:limitMom} & on  \\
 {\tt Ropewalk:pTcut} & 2.0  \\
 {\tt Ropewalk:r0} & 0.41  \\
 {\tt Ropewalk:m0} & 0.2  \\ 
 {\tt Ropewalk:gAmplitud}e & 10.0  \\
 {\tt Ropewalk:gExponent} & 1.0  \\
 {\tt Ropewalk:deltat} & 0.1  \\
 {\tt Ropewalk:tShove} & 1.0  \\
 {\tt Ropewalk:deltay} & 0.1  \\
 {\tt Ropewalk:tInit} & 1.5  \\ [0.5ex] 
 \hline
\end{tabular}
\caption{Input parameters used in Fig.~\ref{fig:plateau} for the shoving model.}
\label{table:shoving}
\end{table}

We also set the two strings' endpoints to have $m_u = 0.33$ GeV, though
this configuration and our massless endpoint configuration
were set to have the same total invariant mass for each string.
Since the shoving model also requires partons to have transverse spacetime
coordinates, we set the strings to be 2.46 fm apart in transverse space
(six times the input
parameter {\tt Ropewalk:r0}). We chose to set the strings relatively far apart,
relative to the transverse radius of the string, since we discovered that for the
above parameter set, a transverse separation between our two straight strings of
$d_{\perp} < 5r_0$ lead to, in our opinion, pathological results.
To understand what each parameter governs in the model, we direct the reader
to \cite{Bierlich:2017vhg}.

\nolinenumbers

\end{document}